%% file: main.tex
\documentclass[journal]{IEEEtran}
 \usepackage{todonotes}
\usepackage{cite}
\usepackage{algpseudocode}
\usepackage{amsmath}
\usepackage{algcompatible}
\usepackage{lscape}
\newtheorem{theorem}{Theorem}

\usepackage{amsfonts,amssymb,amsmath,bbm,mathtools,stackrel,bbm,epsfig}
\usepackage{enumerate}
\usepackage{graphicx}
\usepackage{transparent}
\usepackage{xcolor}
\usepackage{enumitem}
\usepackage[bookmarks,colorlinks]{hyperref}
\usepackage{graphicx,import}

\usepackage{tikz}
\usetikzlibrary{decorations.pathreplacing}
\usepackage{subcaption}
\usepackage[ruled,vlined]{algorithm2e}

\input{myshorts}

\newcommand{\ud}{\,\mathrm{d}}
\usepackage[printonlyused]{acronym}
\acrodef{crb}[CRB]{Cram$\acute{\text{e}}$r-Rao bound}
\acrodef{bcrb}[BCRB]{Bayesian Cram$\acute{\text{e}}$r-Rao bound}
\acrodef{pdf}[PDF]{probability density function}
\acrodef{cdf}[CDF]{cumulative distribution function}
\acrodef{mse}[MSE]{mean-squared-error}
\acrodef{mie}[MIE]{ method of interval estimation}
\acrodef{lmmse}[LMMSE]{linear minimum-mean-squared-error}
\acrodef{mmse}[MMSE]{minimum mean-squared-error}
\acrodef{bcrb}[BCRB]{Bayesian CRB}
\acrodef{wbcrb}[WBCRB]{weighted Bayesian CRB}
\acrodef{snr}[SNR]{signal-to-noise ratio}
\acrodef{wwb}[WWB]{Weiss-Weinstein Bound}
\acrodef{svd}[SVD]{Singular value decomposition}
\acrodef{adc}[ADC]{analog-to-digital converter}
\acrodef{mimo}[MIMO]{multiple-input multiple-output}
\acrodef{fim}[FIM]{Fisher information matrix}
\acrodef{bfim}[BFIM]{Bayesian Fisher information matrix}
\acrodef{iid}[i.i.d.]{independent and identically distributed}
\acrodef{doa}[DOA]{direction of arrival}
\acrodef{scm}[SCM]{sample covariance matrix}
\acrodef{wrt}[w.r.t.]{with respect to}
\acrodef{pmf}[PMF]{probability mass function}
\acrodef{wsn}[WSN]{wireless sensor network}
\acrodef{lgo}[LGO]{linear Gaussian orthonormal}
\acrodef{scm}[SCM]{sample covariance matrices}
\acrodef{mcmc}[MCMC]{Markov Chain Monte Carlo}

\title{Weighted Bayesian Cram$\acute{\text{e}}$r–Rao Bound for Mixed-Resolution Parameter Estimation}
\author{Yaniv Mazor \IEEEmembership{Student Member, IEEE} and
Tirza Routtenberg, \IEEEmembership{Senior Member, IEEE}
\thanks{\footnotesize{Yaniv Mazor and Tirza Routtenberg are with the School of Electrical and Computer Engineering Ben-Gurion University of the Negev Beer-Sheva 84105, Israel, e-mail: \{mazya@post., tirzar@\}bgu.ac.il. This research was supported by the ISRAEL SCIENCE FOUNDATION (Grant No. 1148/22).
 }
} \vspace{-0.25cm}}

\begin{document}
\maketitle
\begin{abstract}
Mixed-resolution architectures, combining high-resolution (analog) data with coarsely quantized (e.g. 1-bit) data, are widely employed in emerging communication and radar systems to reduce hardware costs and power consumption. However, the use of coarsely quantized data introduces non-trivial trade-offs in parameter estimation tasks. In this paper, we investigate 
the derivation of lower bounds for such systems. In particular, we develop 
the weighted Bayesian Cram$\acute{\text{e}}$r–Rao bound (WBCRB) for the mixed-resolution setting with a general weight function. We demonstrate the special cases of: (i) the classical BCRB;  (ii) the WBCRB that is based on the Bayesian Fisher information matrix (BFIM)–Inverse weighting; and (iii) the Aharon-Tabrikian tightest WBCRB with an optimal weight function. Based on the developed WBCRB, we propose a new method to approximate the mean-squared-error (MSE) by partitioning the estimation problem into two regions: (a) where the 1-bit quantized data is informative; and (b) where it is saturated. We apply region-specific WBCRB approximations in these regions to achieve an accurate composite MSE estimate. We derive the bounds and MSE approximation for the linear Gaussian orthonormal (LGO) model, which is commonly used in practical signal processing applications. 
Our simulation results demonstrate the use of the proposed bounds and approximation method in the LGO model with a scalar unknown parameter. It is shown that the WBCRB  outperforms the BCRB, where the BFIM–Inverse weighting version approaches the optimal WBCRB.
Moreover, it is shown that the WBCRB-based MSE approximation is tighter and accurately predicts the non-monotonic behavior of the MSE in the presence of quantization errors. 
\end{abstract}


\section{Introduction}
The estimation of unknown parameters from quantized observations arises in various applications, including \acp{wsn} \cite{RibeiroGiannakis2006P1,Cagatay,xu2022joint,RibeiroGiannakis2006P2}, target tracking \cite{ribeiro2010kalman}, and radar systems \cite{chen2008tracking}. 
In recent years, mixed-resolution architectures, which combine high-resolution (analog) data with coarsely quantized data, such as 1-bit measurements, have been increasingly adopted in modern communication systems, radar, and massive \ac{mimo} technologies {\cite{pirzadeh2018spectral,Xinnan24}}. These architectures offer significant benefits in terms of reduced hardware complexity, power consumption, and bandwidth requirements, making them especially attractive for systems relying on low-cost \acp{adc}.
 For example, communication systems often require energy-efficient and cost-effective \acp{adc}, while still demanding accurate channel estimation \cite{liang2016mixed}. 
 In such settings, parameter estimation is performed using data from multiple resolutions, posing unique challenges. While analog data remains informative across a wide range of \acp{snr}, quantized data exhibits nonlinear behavior and becomes uninformative in high-\ac{snr} regions. 
 This behavior results in a non-monotonic relationship between the \ac{mse} and the \ac{snr}, with a saturation effect at high \ac{snr} where \ac{mse} performance does not improve with increasing \ac{snr} ~\cite{Doaa23,Fesl_Benedikt_2025,Cho24,Stochastic_Resonance,Fu18}; this complicates both estimator design and system analysis.
 Consequently, the development of performance bounds and tractable \ac{mse} expressions for estimation based on mixed analog and quantized data is crucial for reliable performance analysis and system design.

Performance bounds are fundamental tools for the analysis and design of quantized estimation systems.  
Several lower bounds on the non-Bayesian \ac{mse} for deterministic parameter estimation from quantized observations have been proposed, such as \acp{crb} for the linear Gaussian model \cite{Stoica2021,zhu2015parameter,Wang18}, for unknown distributions \cite{RibeiroGiannakis2006P2},  and for additive-controlled perturbation of the quantizer thresholds for $M$-level quantized observations \cite{papadopoulos2001sequential}. 
The \ac{crb} has been developed for different applications, such as for direct position determination
(DPD) \cite{Ni23}.
The \ac{crb} for general mixed-resolution settings was discussed in \cite{harel2017non},
and for  \ac{doa} estimation using a uniform linear array (ULA) in \cite{Xinnan24}. 
In some works, tractable approximations for the \ac{crb} in 1-bit data settings have been derived by replacing the true model with a Gaussian distribution \cite{Sedighi_2021} for nonlinear problems (e.g., \ac{doa} estimation).
In Bayesian settings, \ac{mse} lower bounds for 1-bit  quantized data with dithering have been developed in  \cite{Zeitler_Kramer_Singer_2012,Stephan2023}.
The \ac{bcrb} under
quantized compressed sensing measurements has been derived in \cite{MIMO_OFDM}, and an upper bound on the \ac{bcrb} for the quantized setting was proposed in \cite{Rodrigo,Stein2018}.
The \ac{crb}, hybrid \ac{crb} and \ac{bcrb} for angular-domain channel estimation are all derived in \cite{Liu20}.
MSE bounds have also been used as optimization objectives.  
In \cite{papadopoulos2001sequential}, it is shown that random dithering can significantly reduce the \ac{crb}. In \cite{balkan2010crlb}, deterministic dithering is shown to be optimal in terms of minimizing the  \ac{bcrb}. 
The optimization of the dithering approach based on the \ac{crb} has been explored in \cite{papadopoulos2001sequential}.
Quantization schemes that maximize the \ac{fim} and minimize the CRB were developed in \cite{Aditya_Varshney2014} and \cite{Kar_Varshney2012}, respectively.

While these works provide valuable insights into estimation under quantization, they often lack tightness in high-SNR regimes, or fail to capture the non-monotonic behavior of the MSE of practical estimators, particularly in mixed-resolution settings. For example,  in \cite{RibeiroGiannakis2006P1} it is shown that there is a significant gap between the \ac{crb} for 1-bit data and the \ac{mse} of the clairvoyant estimator.
In particular, in mixed-resolution scenarios the presence of quantized data leads to complex \ac{mse} behavior that is not captured by classical bounds. 
In our previous work \cite{Mazor24}, we demonstrated that the classical \ac{bcrb} fails to capture the non-monotonic behavior of the \ac{mse} in mixed-resolution systems, and thus, cannot be used as a tool for system design.  
This limitation underscores the need for new theoretical tools that account for both analog and quantized 1-bit measurements, provide tight bounds, and enable accurate \ac{mse} approximations across operating regimes.
In this paper, we address this gap by developing the \ac{wbcrb} tailored for mixed-resolution architectures.

In this work, we consider the problem of Bayesian parameter estimation using mixed-resolution data, comprising both analog (unquantized) and 1-bit quantized measurements.
First, we present the implementation of the numerical \ac{mmse} and \ac{lmmse} estimators for a general threshold. Then, we derive the \ac{wbcrb}  tailored for mixed-resolution systems. 
The \ac{wbcrb} is a generalization of the classical \ac{bcrb} that incorporates a user-defined weight matrix to potentially yield tighter bounds. 
We develop three special cases of the \ac{wbcrb}: 1) the classical \ac{bcrb}; 2) 
a data-aware \ac{wbcrb} based on weighting by the inverse of the \ac{fim}; and 3) the tightest \ac{wbcrb}, which is obtained by computing the optimal weight function as proposed in  \cite{Joseph24}.
An additional contribution here is the presentation of a novel two-regime \ac{mse} approximation method that leverages the \ac{wbcrb} to accurately capture the non-monotonic behavior of the \ac{mse} across different \ac{snr} ranges. 
This method partitions the estimation problem into informative and saturated regions of the quantized data and applies region-specific approximations.

We demonstrate the effectiveness of the bounds and approximations for the \ac{lgo} model, a widely used framework in signal processing applications \cite{berman2020resource}.
Our simulation results show that, unlike the \ac{bcrb}, which fails to reflect the \ac{mse} behavior, the proposed \ac{wbcrb} variants accurately track the \ac{mse} and provide tighter alternatives. Moreover, the \ac{wbcrb}-based \ac{mse} approximation closely matches the true \ac{mmse} and \ac{lmmse} performance, making it a practical and computationally efficient tool for system design in resource-constrained mixed-resolution environments.

The remainder of the paper is organized as follows: Section \ref{math} presents mathematically the general mixed-resolution measurement model. Section \ref{est} discusses practical estimation methods. 
In Section~\ref{the_bound}, we derive the general \ac{wbcrb} and outline some special cases. Section \ref{Approximaiton_Method} develops the new regime-aware \ac{mse} approximation method, which leverages the \ac{wbcrb}. In Section \ref{model_LGO}, we apply these results to the \ac{lgo} measurement model. Finally, simulation results are discussed in Section~\ref{sim}, and conclusions are drawn in Section~\ref{conc}.

\textit{Notation:} We use boldface lowercase letters to denote vectors and boldface capital letters for matrices. The identity matrix of size $M\times M$ is denoted by $\Imat_M$, and a vector of ones of length $N$ is denoted by $\onevec_N$. The symbols $(\cdot)^*$,$(\cdot)^T$, and $(\cdot)^H$ represent the conjugate, transpose, and conjugate transpose operators, respectively.  
The notation $\mathbf{A} \succeq \mathbf{B}$ implies that $\mathbf{A} - \mathbf{B}$ is a positive semidefinite matrix. The notation $\text{diag}\left\{\Amat\right\}$ denotes the diagonal matrix containing the diagonal elements of $\Amat$. The $m$th element of the gradient vector $\nabla_{\thetavecsmall} c$ is given by 
$\frac{\partial c}{\partial \theta_m}$,
where $\thetavec = [\theta_1, \dots, \theta_M]^T$ and $c$ is a scalar function of $\thetavec$. 
The notations $\E_p[\cdot]$ and $\E_p[\cdot | A]$ represent the expectation and conditional expectation with respect to a given event $A$, respectively. 
All complex-valued derivatives are defined as the  Wirtinger derivatives \cite[ch. 1]{complex_value}. 
The distribution of a circularly symmetric complex Gaussian random vector with mean $\muvec$ and covariance matrix $\Sigmavec$ is denoted by $\mathcal{CN}(\muvec,\Sigmavec)$, and henceforth referred to as a complex Gaussian vector. We denote the
cross-covariance matrix between $\avec$ and $\bvec$ as $\Cmat_{\avec\bvec}=\E[(\avec-\mu_\avec)(\bvec-\mu_\bvec)^H]$, where $\Cmat_\avec=\Cmat_{\avec\avec}$, and $\mu_\avec$ and $\mu_\bvec$ are the expectations of $\avec$ and $\bvec$, respectively. 
The 1-bit element-wise quantization function is applied separately on the real and imaginary parts of any  $z\in\mathbb{C}$, 
$\Ree\{z\}$ and $\Imm\{z\}$,
and is defined as
\be \label{I.1}
	\mathcal{Q}(z) \hspace{-0.05cm}=\hspace{-0.05cm} \frac{1}{\sqrt{2}} \left[ \begin{cases} 
		\hspace{2.5mm} 1 \:, \Ree\{z\} \geq 0\\
		-1  \:, \Ree\{z\} < 0
	\end{cases} \hspace{-0.35cm} + \hspace{-0.1cm}j\begin{cases} 
		\hspace{2.5mm} 1 \:, \Imm\{z\} \geq 0\\
		-1  \:, \Imm\{z\} < 0
	\end{cases}	\hspace{-3mm}\right].\hspace{-1mm}
\ee

\section{Model and Problem Formulation} \label{math}
In this section, we present the problem of parameter estimation based on mixed-resolution data.
We consider a Bayesian estimation problem where the goal is to estimate a complex-valued continuous random parameter vector, $\thetavec\in\Omega_{\thetavecsmall}\in \mathbb{C}^M$, with the associated prior \ac{pdf} $p(\thetavec)$,
based on mixed-resolution data. 
The available data includes high-resolution analog measurements $\xvec_a \in \mathbb{C}^{N_a}$. This analog data  can be interpreted as a model of sufficiently  high-resolution measurements with negligible residual
quantization noise \cite{liang2016mixed}.
In addition, the data includes a quantized, low-resolution measurement vector, $\xvec_q\in \mathcal{Z}^{N_q}$, which is a random vector with values in the lexicon 
\[{\mathcal{Z}}= \left\{\frac{1}{\sqrt{2}}( 1+ j),\frac{1}{\sqrt{2}}( 1- j),\frac{1}{\sqrt{2}}(- 1+ j),\frac{1}{\sqrt{2}}(-1- j)\right\}.\] This quantized data has been obtained using the 1-bit quantization in \eqref{I.1}.
For the sake of simplicity, we assume that $\xvec_a$  and $\xvec_q$ are conditionally independent given $\thetavec$.

In the following, we describe the joint likelihood function in the presence of both continuous ($\xvec_a$ and $\thetavec$) and discrete ($\xvec_q$) random variables. 
The joint posterior \ac{cdf} of $\xvec_a$ and $\xvec_q$ given $\thetavec$ can be written as
\beqna
\label{joint_cdf}
F_{\xvec_a,\xvec_q|\thetavecsmall}(\alphavec,\betavec|\thetavec)=
\Pr (\xvec_a\leq \alphavec,\xvec_q\leq \betavec|\thetavec) \hspace{1.35cm}\nonumber\\
=
\Pr (\xvec_a\leq \alphavec|\thetavec)
\times \Pr (\xvec_q\leq \betavec|\thetavec),
\eeqna
where $\alphavec \in \mathbb{C}^{N_a}$ and $\betavec \in \mathcal{Z}^{N_q}$.
The last equality stems from the conditional independence of $\xvec_a$ and $\xvec_q$ given $\thetavec$. 

The assumed probability space consists of the sample space \(\mathbb{C}^{N_a}\hspace{-0.1cm}\times \hspace{-0.1cm}\mathcal{Z}^{N_q}\),
the \(\sigma\)-algebra that is the product \(\sigma\)-algebra generated by the Borel sets on \(\mathbb{C}^{N_a}\) and the subsets of \(\mathcal{Z}^{N_q}\), and the probability measure  \(\lambda^{N_a} \times \nu^{N_q}\),  in which $\lambda^{N_a}$ is the Lebesgue measure on $\mathbb{C}^{N_a}$  for the continuous part, and $\nu^{N_q}$ is the counting measure on $\mathcal{Z}^{N_q}$ for the discrete part.   Let\footnote{For simplicity, we denote the \ac{pdf}, the \ac{pmf}, and the mixed likelihood by $p(\cdot)$ (see p. 124 in \cite{morris})} $p(\xvec_a | \thetavec)$ be the \ac{pdf} of $\xvec_a$ 
\ac{wrt} $\lambda^{N_a}$, and $p(\xvec_q | \thetavec)$ be the \ac{pmf} of $\xvec_q$ 
\ac{wrt} $\nu^{N_q}$. Using Bayes theorem, the conditional log-likelihood for the considered model is 
\beqna \label{joint_pdf}
      \log p(\xvec_a|\xvec_q,\thetavec) \hspace{-0.075cm}+\hspace{-0.075cm} \log p(\xvec_q|\thetavec) 
        = \log p(\xvec_a|\thetavec)\hspace{-0.075cm} + \hspace{-0.075cm}\log p(\xvec_q|\thetavec),
\eeqna
where the last equality is since, given $\thetavec$, the measurement vectors $\xvec_a$ and $\xvec_q$ are independent.
This likelihood function is derived using \eqref{joint_cdf} and the Radon-Nikodym theorem \cite{Rudin} \ac{wrt} the probability measure  \(\lambda^{N_a} \times \nu^{N_q}\).


The goal of the considered estimation problem is to use the mixed-resolution measurements,
$\xvec \define [\xvec_a^T, \xvec_q^T]^T$, to estimate $\thetavec$.  
The \ac{mmse} estimator and its associated \ac{mse} do not possess closed-form analytical expressions since the conditional \ac{pdf} of $\thetavec$ given $\xvec$ is intractable in the presence of quantized measurements. 
 Numerical evaluation of the \ac{mmse}  estimator is also impractical for large systems (see Section \ref{est}). Similarly, the  \ac{lmmse} estimator and its \ac{mse}  have analytical expressions 
 only in specific cases, such as in
\cite{pirzadeh2018spectral,wan2020generalized}. Thus,  we aim to develop closed-form performance bounds and tractable \ac{mse} approximations based on the \ac{wbcrb} for performance analysis and system design in the mixed-resolution case.

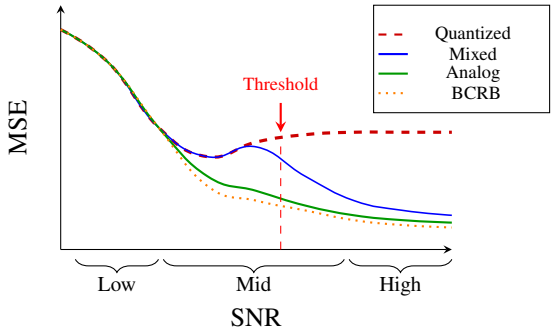
\begin{figure}[h]
\centering
\begin{tikzpicture}[>=stealth,scale=0.65]
    \draw[->] (0,0) -- (8,0);
    \draw[->] (0,0) -- (0,5);
    \node[rotate=90,above] at (-0.5,2.5) {MSE};
    \node[below] at (4,-1.0) {SNR};  

        \draw[very thick,red!80!black,dashed] plot[smooth,tension=0.8] coordinates {
            (0,4.5)    
            (1,3.8)      
            (2,2.5)     
            (3,1.9)
            (4,2.2)    
            (5,2.35)
            (6,2.4)
            (7,2.4)
            (8,2.4)
        };
    \draw[semithick,blue] plot[smooth,tension=0.8] coordinates {
        (0,4.5)    
        (1,3.8)      
        (2,2.5)      
        (3,1.9)    
        (4,2.1)    
        (5,1.5)    
        (6,1.0) 
        (7,0.8)
        (8,0.7)    
    };
    
    \draw[thick,green!60!black] plot[smooth,tension=0.8] coordinates {
        (0,4.5)    
        (1,3.8)      
        (2,2.5)     
        (3,1.5)
        (4,1.2)
        (5,0.9)
        (6,0.7)
        (7,0.6)
        (8,0.55)
    };
    
    \draw[thick,orange,dotted] plot[smooth,tension=0.8] coordinates {
        (0,4.5)    
        (1,3.8)      
        (2,2.5)     
        (3,1.3)
        (4,1.0)
        (5,0.8)
        (6,0.6)
        (7,0.5)
        (8,0.45)
    };
    
    \draw[fill=white,draw=black] (6.4,2.8) rectangle (10,5.0); 
    
    \draw[thick,red!80!black,dashed] (6.5,4.4) -- (7.1,4.4); 
    \node[align=left] at (8.5,4.4) {\scriptsize Quantized };
    
    \draw[thick,blue] (6.5,4.0) -- (7.1,4.0); 
    \node[align=left] at (8.5,4.0) {\scriptsize  Mixed };
    
    \draw[thick,green!60!black] (6.5,3.6) -- (7.1,3.6); 
    \node[align=left] at (8.5,3.6) {\scriptsize Analog };
    
    \draw[thick,orange,dotted] (6.5,3.2) -- (7.1,3.2); 
    \node[align=left] at (8.5,3.2) {\scriptsize BCRB};
    
    \draw [decorate,decoration={brace,amplitude=5pt,mirror},yshift=0pt]
        (0.3,-0.2) -- (2.0,-0.2)
        node[midway,below,yshift=-3pt]{\footnotesize Low};

    \draw [decorate,decoration={brace,amplitude=5pt,mirror},yshift=0pt]
        (2.1,-0.2) -- (5.8,-0.2)
        node[midway,below,yshift=-3pt]{\footnotesize Mid};
        
    \draw [decorate,decoration={brace,amplitude=5pt,mirror},yshift=0pt]
        (5.9,-0.2) -- (8.0,-0.2)
        node[midway,below,yshift=-3pt]{\footnotesize High};

    \node[align=center,red] at (4.5,3.4) {\scriptsize{Threshold}}; 
    \draw[dashed,red] (4.5,0) -- (4.5,2.2); 
    \draw[->,thick,red] (4.5,3.0) -- (4.5,2.4); 
\end{tikzpicture}
\caption{Schematic graph showing \ac{mse} vs. \ac{snr} in a mixed-resolution system with a zero quantization threshold \cite{Mazor24}, showing the three regimes of low-SNR, mid-SNR, and high-SNR. The composite MSE (blue curve) exhibits non-monotonic behavior due to contributions from analog and quantized data. The analog-only MSE (green curve) decreases monotonically, while the quantized-only MSE (red dashed curve) has a distinct peak in the mid-SNR region. The BCRB (orange dotted curve) does not capture the non-monotonicity.}
\label{fig:MSE_schematic}
\end{figure}

Parameter estimation in mixed-resolution systems presents unique challenges due to the non-trivial behavior of the \ac{mse}. 
In the case of purely quantized measurements,  two key regions usually emerge: (1) the low-\ac{snr} regime, where 1-bit quantized measurements convey useful information for the estimation approach; and (2) the high-\ac{snr} regime, where 1-bit quantization saturates, i.e., the measurements can be considered to always obtain a deterministic value \cite{habi2025learnedbayesiancramerraobound,SHAO20,Diana23}.
Similarly, in mixed-resolution systems the \ac{snr} axis can be divided into three regions: 1) low \ac{snr}  regime, where the 1-bit measurements are informative and the analog data is unnecessary; 2)  mid-\ac{snr} regime, where both quantized and analog data contribute to the estimation process; and 3) high \ac{snr}  regime (asymptotic region), occurs at the so-called ``threshold point,”  where the quantized data always get a single, deterministic value in the lexicon space, and thus, only analog data contributes to estimation. 
 The non-monotonic behavior may not be captured by the \ac{bcrb} \cite{Mazor24}, as is illustrated in Fig.~\ref{fig:MSE_schematic} for zero quantization threshold. 
Identifying these regimes through new performance bounds and \ac{mse} approximations is essential for system design and performance analysis.
These tools can inform resource allocation, power allocation, bit-resolution choice, and sensor selection, and serve as benchmarks for practical estimators. 
\vspace{-0.25cm}

\vspace{-0.5cm}
\section{Estimation Methods} \label{est} 
Deriving estimators and calculating their \ac{mse} under the nonlinear mixed-resolution model is analytically demanding.  In this section, we present the background on two benchmark estimators: the \ac{mmse} estimator (Subsection~\ref{mmse_subsection}), which is generally infeasible for large-scale problems, and the more tractable \ac{lmmse} estimator (Subsection~\ref{lmmse_subsection}), which may have a higher \ac{mse}.  Contrasting these two (approximated) estimators
illustrates the intractability of \ac{mse} analysis, and highlights the need for other tools.


\subsection{MMSE Estimator}
\label{mmse_subsection}
The \ac{mmse} estimator of $\thetavec$ given the mixed-resolution observations $\xvec_a$ and $\xvec_q$ is defined as 
\be
\label{mmse_def}
\hat{\thetavec}_{\text{MMSE}}(\xvec_a,\xvec_q) = \E[\thetavec| \xvec_a, \xvec_q].
\ee
Using Bayes rule and the conditional independence between $\xvec_a$ and $\xvec_q$ given $\thetavec$, the posterior \ac{pdf} of $\thetavec$ given the observations is
\be
\label{pdf_theta}
p(\thetavec|\xvec_a, \xvec_q) = \frac{p(\thetavec) \, p(\xvec_a | \thetavec) \, p(\xvec_q | \thetavec)}{ \int_{ \Omega_\thetavecsmall} p(\thetavec) \, p(\xvec_a | \thetavec) \, p(\xvec_q | \thetavec) \, d\thetavec}.
\ee
Substituting \eqref{pdf_theta} in \eqref{mmse_def}, the \ac{mmse} estimator is given by
\be 
\label{mmse_int}
\hat{\thetavec}_{\text{MMSE}}(\xvec_a,\xvec_q) = 
\frac{\int_{\Omega_\thetavecsmall} \thetavec \, p(\thetavec) \, p(\xvec_a | \thetavec) \, p(\xvec_q | \thetavec) \, d\thetavec}{\int_{\Omega_\thetavecsmall} p(\thetavec) \, p(\xvec_a | \thetavec) \, p(\xvec_q | \thetavec) \, d\thetavec}.
\ee
In general, the integrals on the r.h.s. of \eqref{mmse_int} lack closed-form solutions due to the discrete nature of $\xvec_q$ and the nonlinearities in its conditional \ac{pmf}.

Direct numerical evaluation of the integrals in \eqref{mmse_int} faces the curse of dimensionality, as the number of grid points, $M$, grows exponentially with the dimension of $\thetavec$,  which is impractical for high-dimensional problems (see, e.g., \cite{6007031}).
 An alternative is to approximate the \ac{mmse} estimator using Monte Carlo methods \cite{doucet00},\cite[Ch. 3]{Christian00},  which avoids discretization and scales better with $M$. Specifically, \eqref{mmse_int}  can be rewritten as
\be 
\label{mmse_int2}
\hat{\thetavec}_{\text{MMSE}}(\xvec_a,\xvec_q) = 
\frac{\E_\thetavecsmall[\thetavec p(\xvec_a|\thetavec)p(\xvec_q|\thetavec)]}{\E_\thetavecsmall[p(\xvec_a|\thetavec)p(\xvec_q|\thetavec)]}.
\ee 
It should be noted that the expectations in \eqref{mmse_int2} are taken over the prior $p(\thetavec)$ (and not \ac{wrt} the posterior \ac{pdf}), treating $\xvec_a$ and $\xvec_q$ as deterministic vectors. To emphasize this distinction, we use the notation $\E_\thetavecsmall[\cdot]$.

By the Strong Law of Large Numbers, the sample mean converges almost surely to the true expectation. Hence, given $S$ \ac{iid} samples  $\{\thetavec^{(s)}\}_{s=1}^S$ drawn from the prior \ac{pdf} $p(\thetavec)$,  the \ac{mmse} estimator in~\eqref{mmse_int2} can be approximated as
\be
\label{mse_approx}
\hat{\thetavec}_{\text{MMSE}}(\xvec_a,\xvec_q) \approx \frac{\sum_{s=1}^{S} \thetavec^{(s)} p(\xvec_a | \thetavec^{(s)}) p(\xvec_q | \thetavec^{(s)})}{\sum_{s=1}^{S} p(\xvec_a | \thetavec^{(s)}) p(\xvec_q | \thetavec^{(s)})}.
\ee
Unlike grid-based numerical integration, the variances of the empirical estimators in  \eqref{mse_approx} are independent of the dimension $M$ of 
$\thetavec$ \cite{doucet00,Christian00}, making it better-suited for high-dimensional data. However, a large number of samples $S$ is still required for accuracy, especially when $p(\xvec_a | \thetavec)$ and $p(\xvec_q | \thetavec)$ are sharply concentrated.
Advanced techniques, such as importance sampling \cite{Yichuan11}  or \ac{mcmc}, can be employed to reduce the \ac{mse} and accelerate convergence. The overall procedure is summarized in Algorithm~\ref{Algo:MMSE}.

\begin{algorithm} [hbt]
\caption{Monte Carlo MMSE Estimator}
\label{Algo:MMSE}
\begin{algorithmic}[1]
\renewcommand{\algorithmicrequire}{\textbf{Input:}}
\renewcommand{\algorithmicensure}{\textbf{Output:}} 
\REQUIRE ~ \begin{itemize}
    \item Distributions:  $p(\thetavec)$, $p(\xvec_a|\thetavec)$, $p(\xvec_q|\thetavec)$
    \item  Number of samples $S$
    \item Single-test measurements: $\xvec_a$, $\xvec_q$
    	\end{itemize}
\ENSURE  Approximate MMSE estimator $\hat{\thetavec}(\xvec_a,\xvec_q)$.
\FOR {\(s = 1, \dots, S\)}
\STATE Draw the sample $\thetavec^{(s)}\sim p(\thetavec)$
\STATE  Compute the weights 
    $w^{(s)} = p(\xvec_a | \thetavec^{(s)})  p(\xvec_q | \thetavec^{(s)})$
\ENDFOR	
\STATE  Approximate the MMSE in \eqref{mmse_int2} as
  \[
  \hat{\thetavec}(\xvec_a,\xvec_q)= \frac{\sum_{s=1}^{S} \thetavec^{(s)}\, w^{(s)}}{\sum_{s=1}^{S} w^{(s)}}.
  \] 
  \vspace{-0.25cm}
\end{algorithmic} 
\end{algorithm}


Using the law of total expectation, the \ac{mse} matrix of the MMSE estimator  can be written as
\beqna
\label{mse_mat}
{\mathbf{MSE}}(\hat{\thetavec}(\xvec_a,\xvec_q)) \hspace{5cm}
\nonumber\\=
\E\left[ \E\left[ (\hat{\thetavec}(\xvec_a,\xvec_q) - \thetavec)(\hat{\thetavec}(\xvec_a,\xvec_q) - \thetavec)^H \big| \xvec_a, \xvec_q \right] \right],\eeqna
for $\hat{\thetavec}(\xvec_a,\xvec_q)=\hat{\thetavec}_{\text{MMSE}}(\xvec_a,\xvec_q)$.
Numerically evaluating the \ac{mse} matrix in \eqref{mse_mat} is significantly more demanding than computing the \ac{mmse} estimator itself, as it requires an additional expectation over all possible values of $\xvec_a$ and $ \xvec_q$. Specifically, the inner expectation in \eqref{mse_mat} involves computing the posterior covariance matrix, which is highly nonlinear in the presence of discrete variables and non-Gaussian distributions. The outer expectation requires integration over the continuous variable $\xvec_a$ and summation over all discrete lexicon values of $\xvec_q$, where the computational complexity increases as $N_a$ and $N_q$ increase.
These challenges render the direct computation of the Bayesian \ac{mse} matrix impractical in high-dimensional settings, even when employing the Monte Carlo-based approximation that is summarized in Algorithm~\ref{Algo:BayesianMSE}. This empirical \ac{mse} is obtained as the sample mean of the squared error across $K$ trials. The accuracy of this method depends on $ K$ being sufficiently large to ensure convergence.

\vspace{-0.35cm}
\begin{algorithm} [hbt]
\caption{Monte Carlo MSE Matrix}
\label{Algo:BayesianMSE}
\begin{algorithmic}[1]
\renewcommand{\algorithmicrequire}{\textbf{Input:}}
\renewcommand{\algorithmicensure}{\textbf{Output:}} 
\REQUIRE ~ \begin{itemize}
	    \item Distributions: \(p(\thetavec)\), \(p(\xvec_a\mid\thetavec)\), \(p(\xvec_q\mid\thetavec)\)
\item Number of trials \(K\) for MSE evaluation
\item Number of samples \(S\) 
\end{itemize}
\ENSURE  Empirical MSE of the MMSE estimator
\STATE Initialize \(\evec_{\text{total}}=\zerovec\)
\FOR {\(k = 1, \dots, K\)}
\STATE Draw the sample \(\thetavec_0^{(k)} \sim p(\thetavec)\)
\STATE  Generate measurements \(\xvec_a^{(k)}\) and \(\xvec_q^{(k)}\) from 

\(p(\xvec_a \mid \thetavec_0^{(k)})\) and \(p(\xvec_q \mid \thetavec_0^{(k)})\), respectively
\STATE Compute the approximated MMSE estimator \(\hat{\thetavec}^{(k)} = \hat{\thetavec}(\xvec_a^{(k)}, \xvec_q^{(k)})\) using $S$ samples via Algorithm~\ref{Algo:MMSE}
\STATE Update the squared error $\evec_{\text{total}} \leftarrow \evec_{\text{total}} + (\hat{\thetavec}^{(k)} - \thetavec_0^{(k)})(\hat{\thetavec}^{(k)} - \thetavec_0^{(k)})^H$
\ENDFOR	
\STATE  Compute the empirical Bayesian MSE as $
\frac{1}{K}\, \evec_{\text{total}}$.
\end{algorithmic} 
\end{algorithm}

\vspace{-0.75cm}
\subsection{LMMSE Estimator}
\label{lmmse_subsection}
\vspace{-0.05cm}
 The \ac{lmmse} estimator is widely used in scenarios involving quantized data \cite{Benedikt23,Swindlehurst24,Fesl_Benedikt_2025}. While it is generally suboptimal compared to the \ac{mmse} estimator, it often has closed-form expressions or enables tractable numerical approximations, and its computational efficiency makes it well-suited for real-time applications with limited resources \cite{Benedikt23}.

The analytical \ac{lmmse} estimator of $\thetavec$ is 
\be \label{analytical_linear} \hat{\thetavec}_{\text{LMMSE}}(\xvec_a,\xvec_q) = \E[\thetavec]+ \Cmat_{\thetavecsmall \xvec} \Cmat_{\xvec}^{-1} (\xvec-\E[\xvec]), \ee
and its \ac{mse} is given by
\beqna	\label{analytical_lmmse}
{\mathbf{MSE}}\left(\hat{\thetavec}_{\text{LMMSE}}(\xvec_a,\xvec_q)\right) 
= \Sigmavec_\thetavecsmall - \Cmat_{\thetavecsmall \xvec} \Cmat_{\xvec}^{-1} \Cmat_{\thetavecsmall \xvec}^H .
\eeqna
For the mixed-resolution case, $\E[\xvec]=\E[\xvec_a^T,\xvec_q^T]^T$ and
\be	\label{VII.1}
	\Cmat_\xvec = \begin{bmatrix}
		\Cmat_{\xvec_a} & \Cmat_{\xvec_a \xvec_q} \\ \Cmat_{\xvec_q \xvec_a} & \Cmat_{\xvec_q}
	\end{bmatrix},~~
\Cmat_{\thetavecsmall \xvec} = \begin{bmatrix}
	\Cmat_{\thetavecsmall \xvec_a} & \Cmat_{\thetavecsmall \xvec_q}
\end{bmatrix}.	
\ee
In most practical cases, 
the expectations and covariance matrices in \eqref{analytical_linear}--\eqref{analytical_lmmse} do not admit closed-form expressions, and are replaced by sample mean and \ac{scm} estimates based on a dataset of i.i.d. samples, $\mathcal{D} =\{\thetavec^{(k)}, \xvec_a^{(k)}, \xvec_q^{(k)}\}_{k=1}^{K}$.  For example, ${\Cmat}_{\xvec_q}$  is replaced with
\beqna
\hat{\Cmat}_{\xvec_q}=\frac{1}{K} \sum\nolimits_{k=1}^{K} \left(\xvec_q^{(k)}-\bar{\xvec}_q\right)\left(\xvec_q^{(k)}-\bar{\xvec}_q\right)^H,
\eeqna
where $\bar{\xvec}_q = \frac{1}{K} \sum_{k=1}^{K} \xvec_q^{(k)}$.
Replacing the expectations and covariances with empirical estimates yields the numerical \ac{lmmse} estimator, which minimizes the empirical \ac{mse} \cite{shlezinger2023discriminative}
  $\frac{1}{K}\sum_{k=1}^{K} 
       \|\thetavec^{(k)} - \hat{\thetavec}\|_2^2$.
 As  \(K \to \infty\), the \acp{scm} converge to the true covariance matrices (\hspace{-0.02cm}\cite{Serra2014}, p. 728 in \cite{Van}), ensuring asymptotic consistency of this estimator \cite{Serra2014,Van}.


When partial statistical knowledge is available, e.g., under the \ac{lgo} model, one may employ the partially numeric \ac{lmmse} estimator \cite{Mazor24}, which uses known covariance matrices for the analog part and estimates only the covariances involving quantized data. It converges to the \ac{lmmse} estimator as $K$ increases and often outperforms the fully numerical approach. 
Finally, the MSE of the \ac{lmmse} estimator can be computed in two ways: (i) empirically, as in Algorithm 2 by replacing the estimator in Step 4 with \eqref{analytical_linear}, or (ii) directly via \eqref{analytical_lmmse}. Both approaches require \ac{scm} computation, which becomes expensive as $N_a$ and $N_q$ grow. Moreover, since the \ac{lmmse} estimator is suboptimal, its \ac{mse} serves only as an upper bound and may not reflect true system performance.


\vspace{-0.2cm}
\section{WBCRB for mixed-resolution data} \label{the_bound}
\vspace{-0.05cm}
In this section, we develop the \ac{wbcrb} for mixed-resolution observations, which incorporates a flexible weight matrix, enabling tighter and more informative bounds across varying \ac{snr} regimes. 
In Subsection \ref{reg_subsection}, we discuss the associated regularity conditions. In Subsection \ref{bound_subsection}, we present the general bound, and in Subsection \ref{special_cases_subsection}, special cases are discussed.
 We use the general model from Section \ref{math} and a 
 Hermitian, positive-definite weight matrix $\Wmat(\thetavec)\in{\mathbb{C}}^{M\times M}$,  whose entries are differentiable functions, i.e., it has well-defined componentwise Wirtinger derivatives. 

\vspace{-0.25cm}
\subsection{Regularity Conditions}
\label{reg_subsection}
In this subsection, we establish the regularity conditions required for the derivation of the \ac{wbcrb} in the mixed-resolution setting.   Classical formulations of the \ac{bcrb} and \ac{wbcrb} assume continuous observation vectors (see, e.g., \cite{bobrovsky87,Joseph24}), whereas for purely discrete observation vectors,  suitable regularity conditions of the \ac{bcrb} are presented in \cite{Zeitler_Kramer_Singer_2012}. 
 The following \ac{wbcrb} regularity conditions generalize these assumptions to the mixed-resolution model defined in Section~\ref{math}, where the observation vector $\xvec$ comprises both continuous and discrete components, $\xvec_a$ and $\xvec_q$, and the log-likelihood function 
 in \eqref{joint_pdf}.
 The conditions are as follows:
\begin{enumerate} \item\label{cond1}\textbf{Differentiability:} The joint log-likelihood function in \eqref{joint_pdf}
    is differentiable \ac{wrt} $\thetavec$, and the associated \ac{bfim}, $\Jmat$, 
is a well-defined and non-singular matrix. 
\item\label{cond2} \textbf{Integrability:} The score function 
$\nabla_{\thetavecsmall} \log p(\thetavec)p(\xvec_a|\thetavec)p(\xvec_q|\thetavec)$ and its derivatives \ac{wrt} $\thetavec$  are absolutely integrable \ac{wrt} the continuous variables $\thetavec$ and $\xvec_a$, and summable \ac{wrt} the discrete variable $\xvec_q$.
\item\label{cond3}\textbf{Boundary Conditions:}
For each $m\in\{1,\ldots,M\}$, and for every $\xvec_a\in\mathbb{C}^{N_a}$ and 
$\xvec_q\in\mathcal{Z}^{N_q}$, the conditional \ac{pdf} and \ac{pmf} decay fast
enough as $|\theta_m|\rightarrow\infty$ such that integration–by–parts introduces no residual boundary terms in all relevant integrations.
 In particular, we assume that for all indices $l, n, k, m \in \{1, \ldots, M\}$ and all $\xvec_a \in \mathbb{C}^{N_a}$, $\xvec_q \in \mathbb{C}^{N_q}$ and for the Hermitian matrix $\Wmat(\thetavec)\in \mathbb{C}^{M \times M}$, the following limits hold: 
\begin{subequations} \label{bound_con}
    \be \label{con_a}
   \lim_{|\theta_m| \to \infty} \theta_m[\Wmat(\thetavec)]^*_{k,m}p(\thetavec)p(\xvec_a|\thetavec)p(\xvec_q|\thetavec)=0,\\
   \ee
   \be \label{con_b}
   \lim_{|\theta_m| \to \infty} [\Wmat(\thetavec)]^*_{k,m} \frac{\partial [\Wmat(\thetavec)]_{n,l}}{\partial \theta_l^*} p(\thetavec) = 0.
   \ee
\end{subequations}

\item\label{cond4}\textbf{Smoothness:} The conditional expectations of the score functions for the analog and quantized data vanish:
\begin{subequations} \label{cond33}
\be 
\label{16a}
\E_{\xvec_a|\thetavecsmall}[\nabla_{\thetavecsmall} \log p(\xvec_a|\thetavec)|\thetavec]=\zerovec,
\ee
\be 
\label{16b}
\E_{\xvec_q|\thetavecsmall}[\nabla_{\thetavecsmall} \log p(\xvec_q|\thetavec)|\thetavec]=\zerovec.
\ee
\end{subequations}
\end{enumerate}

{\em{Discussion:}} The extension of the \ac{wbcrb} to mixed-resolution observations requires paying attention to the coexistence of continuous and discrete variables. In particular, integrals are replaced with finite summations for the discrete variables $\xvec_q$, as presented e.g., in Condition \ref{cond2}. 
Conditions \ref{cond1}-\ref{cond4} are consistent with those previously established for fully discrete \cite{Zeitler_Kramer_Singer_2012} and continuous models \cite{Joseph24}.
Condition \ref{cond2} allows us to exchange differentiation and expectation via the dominated-convergence theorem.  Formally,
$
\nabla_{\thetavecsmall}\E \bigl[h(\thetavec,\xvec)\bigr]
         = \E\bigl[\nabla_{\thetavecsmall} h(\thetavec,\xvec)\bigr],
$
for any measurable function  $h(\thetavec,\xvec)$, where the outer expectation (\ac{wrt} $\xvec_q$) is a finite sum in this case.

For the BCRB (obtained for $\Wmat(\thetavec)=\Imat$) and continuous observation only, Conditions~\eqref{con_a}–\eqref{con_b} reduce to the familiar requirement that the joint \ac{pdf} vanishes at the edges, 
$\lim_{\theta_m \to \pm \infty}\theta_m p(\thetavec,\xvec_a)=0$ (see, e.g., \cite[Eq. (4.178)]{van_bell}).  In the weighted case, an extra decay condition, \eqref{con_b}, is necessary because the integration–by–parts step multiplies the \ac{pdf} and \ac{pmf} by $\Wmat(\thetavec)$ and its derivatives. 
For Condition \ref{cond4}, the continuous part in \eqref{16a} is automatic when the \ac{pdf} is differentiable and integrable, if the support of $p(\xvec_a|\thetavec)$ is independent of $\thetavec$ \cite{van_bell}.  For the mixed-resolution case, \eqref{16b}  must be stated explicitly to guarantee that the \ac{bfim} decomposes to analog, quantized, and prior contributions, as presented in the following.

\subsection{General WBCRB}
\label{bound_subsection}
We now present the main theorem of the \ac{wbcrb}. To this end, we first develop the \ac{bfim} term. 
Under Conditions \ref{cond1} and \ref{cond2} from Section \ref{reg_subsection}, the \ac{bfim} for our model is well-defined.
We use the fact that the \ac{bfim} satisfies \cite[p. 173]{complex_value}
\be	\label{Fisher_information}
	\Jmat = 
    \Jmat_\thetavecsmall+\E_\thetavecsmall[\Jmat_{\xvec|\thetavecsmall}(\thetavec)],
\ee 
where the prior \ac{fim} is defined as
\be \label{FIMp2}
	\Jmat_\thetavecsmall\triangleq
\E_\thetavecsmall[\nabla_{\thetavecsmall}^{H}\log p(\thetavec)                                \,\nabla_{\thetavecsmall}\log p(\thetavec)]
\ee
and $\Jmat_{\xvec|\thetavecsmall}(\thetavec)$ is the posterior \ac{fim}.
In addition, using \eqref{joint_pdf} and \eqref{cond33}, the posterior \ac{fim} is given by
\be \label{Fisher_observ}
{\rm{E}}_\thetavecsmall[\Jmat_{\xvec|\thetavecsmall}(\thetavec)] = {\rm{E}}_\thetavecsmall[\Jmat_{\xvec_a|\thetavecsmall}(\thetavec)]+{\rm{E}}_\thetavecsmall[\Jmat_{\xvec_q|\thetavecsmall}(\thetavec)],
\ee
where the \ac{fim}s based on each set of measurements $\xvec_a$ and $\xvec_q$, respectively, given $\thetavec$, are (see, e.g., \cite{Zeitler_Kramer_Singer_2012,kay1993fundamentals})
\begin{subequations} \label{FIMdata}
\beqna
	\label{FIMa}
	\Jmat_{\xvec_a|\thetavecsmall}(\thetavec)\define \E_{\xvec_a|\thetavecsmall}\left[\nabla_{\thetavecsmall}^H \log 
 {p}(\xvec_a|\thetavec) \nabla_{\thetavecsmall} \log  {p}(\xvec_a|\thetavec)|\thetavec\right],
\\	\label{FIMq_i}
	\Jmat_{\xvec_q|\thetavecsmall}(\thetavec) \define \E_{\xvec_q|\thetavecsmall}\left[\nabla_{\thetavecsmall}^H\log {p}(\xvec_q|\thetavec)\nabla_{\thetavecsmall}\log {p}(\xvec_q|\thetavec)|\thetavec\right].
\eeqna
\end{subequations}
By substituting these definitions in \eqref{Fisher_information}, we obtain the following structure of the \ac{bfim} for this case:
\be	\label{J_def}
	\Jmat = \Jmat_\thetavecsmall+\E_\thetavecsmall[\Jmat_{\xvec_a|\thetavecsmall}(\thetavec)]+\E_\thetavecsmall[\Jmat_{\xvec_q|\thetavecsmall}(\thetavec)].
\ee 

In addition, we define the following terms:
\be
\label{v_def}
v_k(\thetavec) \define\sum\nolimits_{m=1}^M\frac{\partial[\Wmat(\thetavec)]_{k,m}}{\partial\theta_m^*}, ~k=1,\ldots,M,
\ee
\be \label{A_def}
[\Amat]_{n,k}\define -\E\left[\sum\nolimits_{m=1}^M \frac{\partial  \left([\Wmat(\thetavec)]_{n,m}v^*_k(\thetavec)\right)}{\partial \theta_m^*}\right],
\ee
$k,n=1,\ldots,M$, and
\beqna \label{Gmat}
\Gmat = \E\big[ \Wmat(\thetavec) \nabla_{\thetavecsmall}^H \log p(\thetavec)p(\xvec_a|\thetavec)p(\xvec_q|\thetavec) \hspace{0.5cm}
\nonumber\\
\times
\nabla_{\thetavecsmall} \log p(\thetavec)p(\xvec_a|\thetavec)p(\xvec_q|\thetavec) \Wmat^H(\thetavec) \big] \nonumber\\
+\Amat+\Amat^H+\E[\vvec(\thetavec)\vvec^H(\thetavec)].\hspace{1.5cm}
\eeqna
It can be seen that the matrix $\Gmat$ is a positive-definite matrix, where, except for the first term and the computation of the expectations,  all of its components depend only on the chosen weight matrix $\Wmat(\thetavec)$ and its derivatives.
\begin{theorem}[Mixed-resolution WBCRB] \label{theo_1}
Assume the model 
 stated in Section \ref{math} with
$\thetavec \in \Omega_\thetavecsmall \subset \mathbb{C}^M$, $\xvec_a \in \mathbb{C}^{N_a}$, and $\xvec_q \in \mathcal{Z}^{N_q}$, and where $\xvec_a$ and $\xvec_q$ are conditionally independent given $\thetavec$.
In addition,  suppose that Conditions \ref{cond1}-\ref{cond4} hold. 
Then, the \ac{mse} matrix of any estimator $\hat{\thetavec}(\xvec_a, \xvec_q)$ satisfies
\be \label{bound_wbcrb}
\E \left[ (\hat{\thetavec}(\xvec_a, \xvec_q) - \thetavec)(\hat{\thetavec}(\xvec_a, \xvec_q) - \thetavec)^H \right] \succeq {\text{WBCRB}},
\ee
where 
\be \label{wbcrb_def}
{\text{WBCRB}}\define \E[{\Wmat}(\thetavec)]\Gmat^{-1}\E[{\Wmat}^H(\thetavec)],
\ee
in which $\Gmat$ is defined in \eqref{Gmat}.
\end{theorem}
\begin{IEEEproof}
The proof appears in Appendix \ref{AppA}.
\end{IEEEproof}
In the supplemental material attached to this paper, we discuss the extension of this theorem to a widely-linear lower bound on the \ac{mse} of the augmented complex-valued error vector.

\subsection{Special Cases}
\label{special_cases_subsection}
    We next examine several choices of the weighting matrix~\(\Wmat(\thetavec)\), each yielding a particular bound of the general \ac{wbcrb} form of Theorem~\ref{theo_1}.  

\subsubsection{BCRB} \label{special_bcrb}
By choosing $\Wmat(\thetavec)=\Imat$, it can be seen from \eqref{v_def} and \eqref{A_def} that we get $\vvec(\thetavec)=\zerovec$ and $\Amat=\zerovec$.  Thus, \eqref{Gmat} implies that in this case
\beqna 
\Gmat = \E\big[ \nabla_{\thetavecsmall}^H \log p(\thetavec)p(\xvec_a|\thetavec)p(\xvec_q|\thetavec) \hspace{0.75cm}
\nonumber\\
\times
\nabla_{\thetavecsmall} \log p(\thetavec)p(\xvec_a|\thetavec)p(\xvec_q|\thetavec)  \big]=
	\Jmat,  
\eeqna
where $\Jmat$ is the \ac{bfim} defined in \eqref{J_def}.
Thus, the \ac{wbcrb} from \eqref{wbcrb_def} in this case is reduced to 
\be \label{bcrb}
{\text{WBCRB}}= \Jmat^{-1}={\text{BCRB}},
\ee
which is the classical \ac{bcrb}. The \ac{bcrb} under the \ac{lgo} model is discussed in \cite{Mazor24} and in Section \ref{model_LGO}.
\subsubsection{\ac{fim}–Inverse Weighting} \label{case_at}
A common data–aware weighting choice is the inverse of the internal term of the \ac{bfim} before computing the expectation \ac{wrt} $\thetavec$. That is, to take 
 \beqna \label{w_choice} \Wmat(\thetavec)=\tilde{\Jmat}^{-1}(\thetavec),
 \eeqna
 where
 \beqna
 \label{J_tilde_def}
 \tilde{\Jmat}(\thetavec)
 \define
 \nabla_{\thetavecsmall}^{H}\log p(\thetavec)                                \,\nabla_{\thetavecsmall}\log p(\thetavec)
+\Jmat_{\xvec_a|\thetavecsmall}(\thetavec)+\Jmat_{\xvec_q|\thetavecsmall}(\thetavec).
 \eeqna
 It can be seen that $ \E_\thetavecsmall[\Wmat^{-1}(\thetavec)]=\E_\thetavecsmall[\tilde{\Jmat}(\thetavec)]=\Jmat$, where $\Jmat$ is defined in \eqref{Fisher_information}.
By substituting \eqref{w_choice} into \eqref{Gmat},
 we get
\be
\label{at-bcrb}
\Gmat = \E\big[\Wmat(\thetavec)\big]
+\Amat+\Amat^H+\E\big[\vvec(\thetavec)\vvec^H(\thetavec)\big],
\ee
that together with \eqref{w_choice}  defined the WBCRB in  \eqref{wbcrb_def} for this choice of weighting matrix $\Wmat(\thetavec)$.
In the simulations, we demonstrate that this choice offers both analytical tractability and competitive performance across the tested scenarios.

The equality in \eqref{bound_wbcrb} holds (i.e., the bound is tight) {\em{iff}}  $\Jmat_{\xvec|\thetavecsmall}$ defined in \eqref{FIMdata} is independent of $\thetavec$ \cite{Joseph24}. 
For example, in the pure analog Gaussian case in a linear additive model, this condition is satisfied, and choosing $\Wmat(\thetavec) = \Jmat_{\xvec_a|\thetavecsmall}^{-1}$ yields a tight bound  (also, in this case the \ac{wbcrb} coincides with the conventional \ac{bcrb} as well).



For the matrix $\Gmat$,  the first term in \eqref{at-bcrb} reflects the chosen weight matrix $\Wmat(\thetavec)$, while the remaining three terms depend on the derivatives of the weight matrix.  
If these derivatives (i.e., $|v_k(\thetavec)| $, $k=1,\ldots,M$) are significantly smaller than the elements of $\Wmat(\thetavec)$ from \eqref{w_choice}, then these three terms are negligible and \eqref{at-bcrb} is reduced to
$\Gmat \approx \E\big[\Wmat(\thetavec)\big]$.
In addition, in the asymptotic case of large $N_a,N_q$, the prior information is neglected.
In this case, the bound defined by  \eqref{wbcrb_def}, \eqref{w_choice}, and \eqref{at-bcrb} can be computed using expectation \ac{wrt} the prior as
\be \label{wbcrb_ecrb}
{\text{WBCRB}} \approx \E_\thetavecsmall [(\Jmat_{\xvec_a|\thetavecsmall}(\thetavec)+\Jmat_{\xvec_q|\thetavecsmall}(\thetavec))^{-1}]={\text{ECRB}}.
\ee
The r.h.s. of \eqref{wbcrb_ecrb}
is the expected \ac{crb} \cite[eq. 39]{van_bell} for mixed-resolution data.
Using Jensen's inequality,  \eqref{wbcrb_ecrb}  implies that for a high number of measurements $N_a+N_q$, we get
\be
{\text{ECRB}} \succeq {\text{BCRB}}.
\ee
Thus, under \eqref{wbcrb_ecrb}, the \ac{wbcrb} appears to yield a tighter bound than the \ac{bcrb} 
as long as \ac{bfim} is non-singular. 

The term in \eqref{wbcrb_ecrb} is valid for an asymptotic number of observations,
but not for an asymptotic \ac{snr}. In particular, for pure 1-bit systems, the data \ac{bfim} $\Jmat_{\xvec_q|\thetavecsmall}(\thetavec)$ is a highly nonlinear function of $\thetavec$, and the derivative terms in $\Gmat$ generally do not vanish for a finite number of measurements.
In contrast, in pure analog systems or in mixed-resolution systems where the analog observations dominate as the \ac{snr} increases, the chosen weight matrix from \eqref{w_choice} is typically smooth and well-behaved, 
leading to vanishing derivative terms and a valid reduction of the \ac{wbcrb} to the expected \ac{crb} as in \eqref{wbcrb_ecrb}. 
In general, even in the non-asymptotic case, the \ac{wbcrb} with $\Wmat(\thetavec)$ from \eqref{w_choice} offers a tighter and more reliable performance bound than the \ac{bcrb}, as shown in the simulations.

\subsubsection{Optimal WBCRB} \label{sec_opt}
For the sake of simplicity, we present the optimal \ac{wbcrb} for the scalar case ($M=1$), as derived in \cite{Joseph24}.
We consider a finite parameter space $\Omega_{\thetavecsmall}$ and some test points, 
$\theta_1, \dots, \theta_L$, which are obtained from sampled parameters over $\Omega_{\thetavecsmall}$ with $\Delta$ spacing. Then, we select the weighting vector (instead of the matrix $\Wmat(\thetavec)$, since we are estimating a scalar), as  
\be \label{opt}
    \wvec(\thetavec) = \frac{\left( \mathbf{ZF} + \bar{\mathbf{\Psi}} \right)^{-1} \fvec}{\fvec^T \left( \mathbf{ZF} + \bar{\mathbf{\Psi}} \right)^{-1} \fvec},
\ee
where 
$\fvec = \Delta \cdot [p_{\theta}(\theta_1), \dots p_{\theta}(\theta_L)]^T $,    
$\mathbf{F} = \text{diag}(\fvec)$,  
$\Zmat = \text{diag} (\tilde{J}(\theta_1), \dots, \tilde{J}(\theta_L))$,  in which $\tilde{J}(\theta)$ is defined in \eqref{J_tilde_def}. In addition,
$\Kmat$ is a discrete-derivative matrix:  
\be
\Kmat_{nm} = \frac{1}{\Delta} 
\begin{cases} 
1 & n = m \\ 
-1 & n = m + 1 \\ 
0 & \text{otherwise}
\end{cases},
\ee
and 
$\overline{\mathbf{\Psi}}= -\left( \mathbf{FKK} + (\mathbf{FKK})^T + \mathbf{K}^T \mathbf{FK} \right)$. 
Substituting \eqref{opt} in \eqref{wbcrb_def}, results in the tightest \ac{wbcrb} \cite{Joseph24}:
\be \label{opt_wbcrb}
{\text{WBCRB}} = \fvec^T\left( \mathbf{ZF} + \bar{\mathbf{\Psi}} \right)^{-1} \fvec.
\ee


\section{MSE Approximation Method} \label{Approximaiton_Method} 
In nonlinear Bayesian estimation with quantized measurements, classical lower bounds, such as the \ac{bcrb},  often fail to capture the non-monotonic behavior of the 
\ac{mse} across the \ac{snr} range, as illustrated in Fig.~\ref{fig:MSE_schematic} and discussed in \cite{Mazor24}.
This issue is especially pronounced in mixed-resolution systems. At low and moderate \ac{snr} levels, quantized observations provide useful information; however, at high \ac{snr} levels, the quantized outputs saturate, i.e., converge to constant values (e.g., all-ones), and convey only coarse region or sign-level details \cite{mezghani2007ultra, bar2002doa,li2017channel}.
To address this, we propose a practical \ac{mse} approximation method that leverages the \ac{wbcrb} derived in Section~\ref{the_bound}, while avoiding computationally intensive  Monte Carlo simulations of the estimator. The method exploits the regime-switching behavior of quantized data by combining the \ac{wbcrb} (in the informative regime) with an analog-only \ac{mse} approximation (in the saturation regime). This approach is reminiscent of the \ac{mie}, which approximates performance under threshold effects by partitioning the parameter space into informative and non-informative regions \cite[ch.~4.4.2]{van2004detection}, \cite{Athley02}.

The non-monotonic behavior arises because, as the \ac{snr} increases, the noise becomes negligible
and the quantized output  $\xvec_q$ converges to a deterministic function of $\thetavec$.
 In this regime, repeated observations yield identical quantized values, and the posterior distribution is effectively determined solely by the prior and any analog measurements, with negligible impact of $\xvec_q$.
Formally, this property can be written as
\be
\label{eq:asymptotic_pmf}
p(\xvec_q|\thetavec)
\overset{\text{SNR} \rightarrow \infty}
{\longrightarrow}
\begin{cases}
    1 & \xvec_q=\cvec(\thetavec)\\
    0& {\text{otherwise}}
\end{cases}, 
\ee
where $\cvec(\thetavec)$  is a deterministic mapping from the parameter space to the quantized alphabet, independent of the noise realization 
\cite{Fu18},\cite[p. 112]{Stochastic_Resonance}. 
We define the \emph{decision cell} (or quantization cell) associated with the observed $\xvec_q$ as
\begin{equation}
\label{eq:decision_cell}
\mathcal{S}(\xvec_q) \triangleq \{\thetavec \in \Omega_{\thetavecsmall} \; : \; \cvec(\thetavec) = \xvec_q\},
\end{equation}
i.e., the set of parameter values producing the same quantized output at high-\ac{snr}.
Substituting \eqref{eq:asymptotic_pmf} into \eqref{pdf_theta} yields 
\begin{equation}
\label{eq:posterior_cell}
p(\thetavec |\xvec_a, \xvec_q) \overset{\text{SNR} \rightarrow \infty}{\longrightarrow}
\frac{p(\thetavec)\,p(\xvec_a |\thetavec)\,\mathbbm{1}_{\{\thetavecsmall \in \mathcal{S}(\xvec_q)\}}}
{ \int_{\mathcal{S}(\xvec_q)} p(\thetavec')\,p(\xvec_a | \thetavec')\,d\thetavec'},
\end{equation}
where $\mathbbm{1}_{A}$ denotes the indicator of the event A.
Thus, at high \ac{snr} levels, $\xvec_q$ simply restricts the posterior support to $\mathcal{S}(\xvec_q)$; within this set, the posterior shape is governed solely by the prior $p(\thetavec)$ and the analog likelihood $p(\xvec_a | \thetavec)$.
In particular, if the analog data is absent, the posterior reduces to the prior truncated to $\mathcal{S}(\xvec_q)$ and renormalized.

To characterize the non-monotonic \ac{mse} behavior in mixed-resolution systems, we partition the observation space into informative and saturation regimes. The latter is  the event
$\mathcal{N} $ in which \eqref{eq:asymptotic_pmf} holds,
i.e., the signal lies in one cell. 
 Under $\mathcal{N}$, $\xvec_q$ becomes conditionally deterministic 
and conveys negligible information about $\thetavec$. Thus, under $\mathcal{N} $, the estimation performance is dominated by the analog measurements.
Using the law of total expectation, the \ac{mse} can be decomposed as
\beqna \label{appro}
\text{MSE}(\hat{\thetavec}) =
\E \left[
(\hat{\thetavec} - \thetavec)(\hat{\thetavec}- \thetavec)^H \,|\,\mathcal{N}
\right] 
\cdot \Pr(\mathcal{N}) \hspace{1.25cm}\nonumber\\ 
+\E \left[
(\hat{\thetavec} - \thetavec)(\hat{\thetavec}- \thetavec)^H \,|\, \mathcal{N}^c
\right] 
\cdot (1-\Pr(\mathcal{N})),
\eeqna
where $\mathcal{N}^c$ denotes the complementary event of $\mathcal{N}$ and $\Pr(\mathcal{N})$ can be evaluated as
\begin{equation} \label{prob_n}
\Pr(\mathcal{N}) = \mathbb{E}_{\thetavecsmall}\!\left[ \Pr(\mathcal{N} | \thetavec) \right].
\end{equation}

Under the reasonable assumption that $\xvec_a$ is informative enough, the indicator $\mathbbm{1}_{\{\thetavecsmall \in \mathcal{S}(\xvec_q)\}}$  in \eqref{eq:asymptotic_pmf} is redundant, and~\eqref{eq:posterior_cell} can be approximated via the analog-only posterior $p(\thetavec|\xvec_a)$, i.e., 
$
     p(\thetavec | \xvec_a, \mathcal{N}) \overset{\text{SNR} \rightarrow \infty}{\longrightarrow} p(\thetavec | \xvec_a).$
   Consequently, the \ac{mmse} estimator based on $(\xvec_a, \xvec_q)$ under $\mathcal{N}$ approximately coincides with the \ac{mmse} estimator based only on $\xvec_a$, i.e.,
    \[
    \hat{\thetavec}_{\text{MMSE}}(\xvec_a, \xvec_q) \approx \hat{\thetavec}_{\text{MMSE}}(\xvec_a), \quad \text{under } \mathcal{N},
    \]
    where $\hat{\thetavec}_{\text{MMSE}}(\xvec_a)$ is the \ac{mmse} estimator based solely on the analog data.
Consequently, similar to \eqref{mse_mat}, 
for the \ac{mmse} estimator we have
\beqna
\label{45_eq}
\E\left[ \E\left[ (\hat{\thetavec}_{\text{MMSE}}(\xvec_a,\xvec_q) - \thetavec)
\right.\right. \hspace{4cm} \nonumber\\
\left. \left. \times (\hat{\thetavec}_{\text{MMSE}}(\xvec_a,\xvec_q) - \thetavec)^H \big| \xvec_a, \xvec_q, \mathcal{N}\right] |\mathcal{N}\right]\hspace{2cm}
\nonumber\\  \approx
\E\left[ \E\left[ (\hat{\thetavec}_{\text{MMSE}}(\xvec_a) - \thetavec)(\hat{\thetavec}_{\text{MMSE}}(\xvec_a) - \thetavec)^H \big| \xvec_a \right] \right].
\eeqna

To maintain analytical tractability, we further approximate the \ac{mmse} \ac{mse} in \eqref{45_eq} using the closed-form \ac{mse} of the \ac{lmmse} estimator:
\beqna
\label{approx_under_N}
\E\left[ (\hat{\thetavec}_{\text{MMSE}}(\xvec_a) - \thetavec)(\hat{\thetavec}_{\text{MMSE}}(\xvec_a) - \thetavec)^H  \right] 
\nonumber\\
\approx {\mathbf{MSE}}(\hat{\thetavec}_{\text{LMMSE}}(\xvec_a)), 
\eeqna
where, similar to 
\eqref{analytical_lmmse}-\eqref{VII.1}, the \ac{mse} of the analog-only \ac{lmmse} estimator  is 
\beqna
\label{MSE_LMMSE_analog}
{\mathbf{MSE}}(\hat{\thetavec}_{\text{LMMSE}}(\xvec_a))
= \Sigmavec_\thetavecsmall - \Cmat_{\thetavecsmall \xvec_a} \Cmat_{\xvec_a}^{-1} \Cmat_{\thetavecsmall \xvec_a}^H .
\eeqna
If the analog data is absent, the posterior in \eqref{eq:posterior_cell} is reduced to
\begin{equation}
\label{eq:posterior_cell_1_bit}
p(\thetavec | \xvec_q) \overset{\text{SNR} \rightarrow \infty}{\longrightarrow}
\frac{p(\thetavec)\,\mathbbm{1}_{\{\thetavecsmall \in \mathcal{S}(\xvec_q)\}}}
{ \int_{\mathcal{S}(\xvec_q)} p(\thetavec')\,d\thetavec'}.
\end{equation}
Thus, instead of \eqref{MSE_LMMSE_analog},  the \ac{mse} of the \ac{mmse} estimator that is obtained from the posterior in \eqref{eq:posterior_cell_1_bit} should be used.

In the informative region $\mathcal{N}^c$, the quantized observations have non-negligible variability, and 
 we approximate the conditional \ac{mse} via the mixed-resolution \ac{wbcrb}, as follows:
\begin{equation}
\label{approx_under_N_c}
\E[(\hat{\thetavec} - \thetavec)(\hat{\thetavec} - \thetavec)^H \,|\, \mathcal{N}^c] \approx \text{WBCRB}.
\end{equation}
The choice of weighting matrix in the \ac{wbcrb} can be problem-specific; we adopt the selection in~\eqref{w_choice}, which yields accuracy close to the optimal choice in our tests (Section~\ref{sim}).

By substituting the two approximations from \eqref{approx_under_N} and \eqref{approx_under_N_c}
in \eqref{appro}, we obtain
\beqna \label{analog_app}
\text{MSE}(\hat{\thetavec}) 
\approx
{\mathbf{MSE}}(\hat{\thetavec}_{\text{LMMSE}}(\xvec_a)) 
\cdot \Pr(\mathcal{N}) \nonumber\\+\text{WBCRB}
\cdot (1-\Pr(\mathcal{N})).
\eeqna
This approximation captures the transition in informativeness across the \ac{snr} range.

In general, the probability $\Pr({\mathcal{N}})$ in \eqref{analog_app} does not have closed-form expressions. 
We therefore adopt a lightweight Monte Carlo approach, similar to that used in Subsection~\ref{lmmse_subsection}, as summarized in Algorithm~\ref{Algo:MC_region}.
The computation in Algorithm~\ref{Algo:MC_region} constitutes the main computational component of the proposed \ac{mse} approximation, and its complexity is dominated by data generation and equality checking of $S$ $\xvec_q$ samples, resulting in a per-sample cost of $\mathcal{O}(N_qS)$. This procedure is considerably more efficient than estimating the \ac{mse} of the \ac{mmse} estimator using Monte Carlo integration (see Algorithm~\ref{Algo:MMSE}), which requires computing high-dimensional conditional expectations for each sample and averaging the error.  In addition, empirical results indicate that a modest number of iterations  (e.g., $K = 10^2$--$10^3$) suffices for stable estimation of $\Pr(\mathcal{N} | \thetavec) $. 
Finally, for certain models, $\Pr(\mathcal{N}|\thetavec)$ admits a closed-form expression (see Subsection \ref{MSE_approx_LGO}).
\vspace{-0.25cm}
\begin{algorithm}[hbt]
\caption{Monte Carlo Saturation Probability}
\label{Algo:MC_region}
\begin{algorithmic}[1]
\renewcommand{\algorithmicrequire}{\textbf{Input:}}
\renewcommand{\algorithmicensure}{\textbf{Output:}} 
\REQUIRE ~
\begin{itemize}
    \item Distributions:  $p(\thetavec)$, $p(\xvec_q|\thetavec)$
    \item Number of iterations: \( K \)
    \item Number of samples: \( S \)
\end{itemize}
\ENSURE Estimated probability \( \widehat{\Pr}(\mathcal{N}) \)
\STATE Initialize counter: \( c_{\text{sat.}} \gets 0 \)
\FOR{\(k = 1, \dots, K\)}
    \STATE Draw the sample \( \thetavec^{(k)} \sim p(\thetavec) \)
    \STATE Generate $S$ i.i.d samples ${\xvec_q}^{(k,s)} \sim p(\xvec_q | \thetavec^{(k)}) $
  \STATE \If{$\mathbf{x}_q^{(k,s-1)} = \mathbf{x}_q^{(k,s)}$ for all $s=2,\dots,S$}{
    $c_{\text{sat.}} \leftarrow c_{\text{sat.}} + 1$\;
  }
\ENDFOR
\STATE \( \widehat{\Pr}(\mathcal{N}) = c_{\text{sat.}} / K. \)
\end{algorithmic}
\end{algorithm}



\vspace{-0.25cm}
\section{LGO Model} \label{model_LGO}
\vspace{-0.05cm}
In this section,  we discuss the special case of parameter estimation with mixed-resolution data under the widely-used \ac{lgo} model. 
In Subsection \ref{model_subsec} we present this model.  
The \ac{wbcrb} from Section \ref{the_bound} is developed for the \ac{lgo} model in Subsection \ref{WBCRB_LGO}, and in Subsection \ref{MSE_approx_LGO} we demonstrate the \ac{mse} approximation from Section \ref{Approximaiton_Method}. 
\subsection{The LGO Model} \label{model_subsec}
\begin{figure}[hbt]
	\centering	\includegraphics[width=0.9\linewidth]{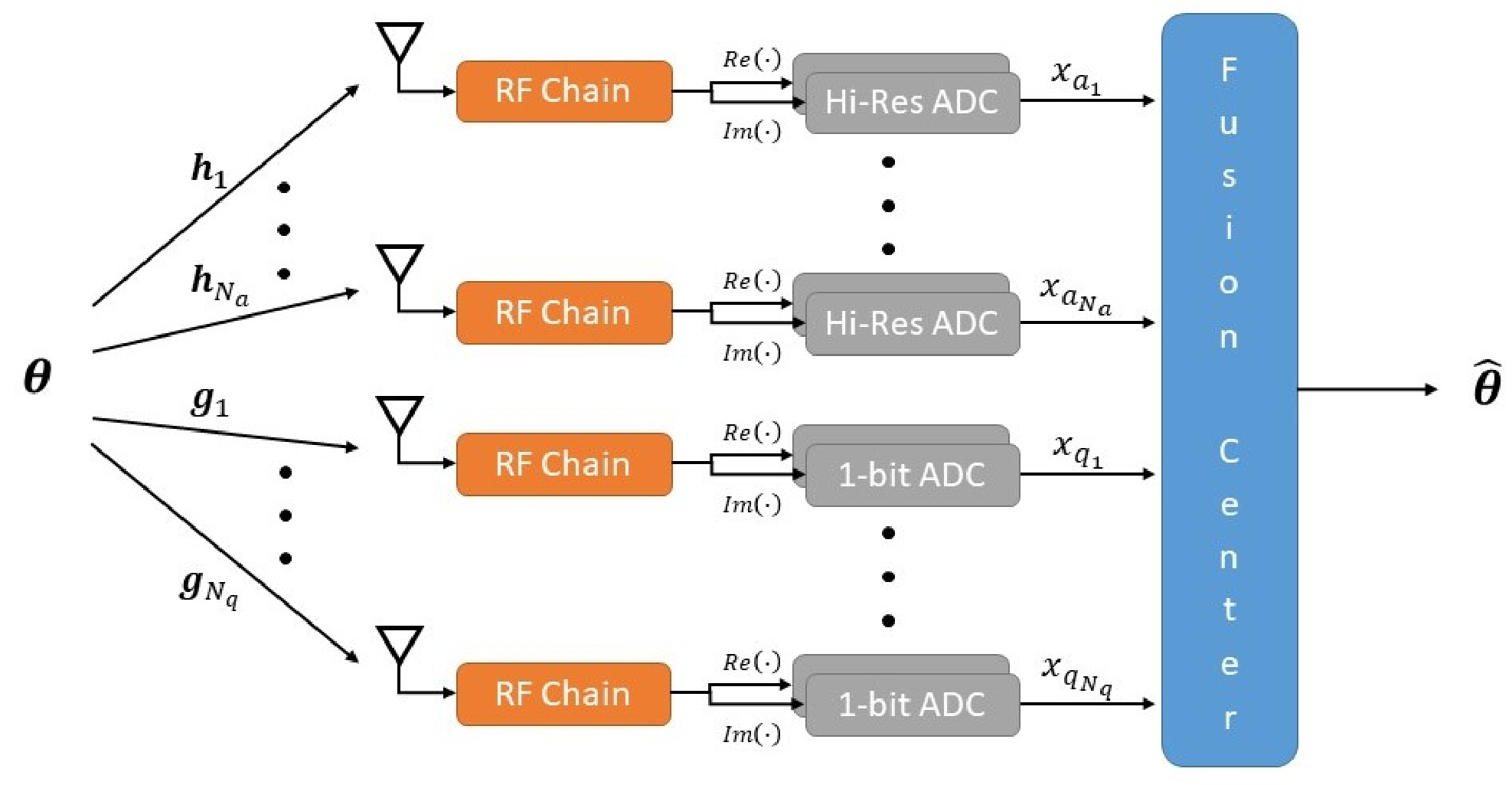}
	\caption{Channel estimation with mixed-resolution measurements: a channel, $\thetavec$, is estimated by transmitting a pilot sequence that is received by either high- or 1-bit low-resolution quantization. The unknown channel is estimated from the mixed-resolution measurements, $\xvec_{a}$, $\xvec_{q}$, in the fusion center.}
	\label{II.A.Fig1}
\end{figure}

In our previous work in \cite{berman2020resource}, we showed that the following significant problems are special cases
of the \ac{lgo} model:
\begin{itemize}
\item Channel estimation in \ac{mimo} communication systems \cite{park2017optimization,berman2020resource,berman2021partially,li2017channel,Swindlehurst14} (see Section V of \cite{berman2020resource}). An illustration of the \ac{mimo} system is shown in Fig. \ref{II.A.Fig1}, where the parameter vector, $\thetavec$, is estimated at the fusion center using the mixed-resolution data under the \ac{lgo} model.
    \item Scalar parameter estimation, which is widely used in \ac{wsn}s  \cite{RibeiroGiannakis2006P1,papadopoulos2001sequential} (see Section IV 
 in \cite{berman2020resource}).
\item Sequential linear models, where measurements are taken over time with the same system matrices, but with possibly different \ac{adc} resolutions (see
Section V.A of \cite{berman2020resource}).
\end{itemize}
Tractable bounds can, hence, be used for these commonly used applications. 

Mathematically, the \ac{lgo} model assumes:
\begin{itemize}
    \item The parameter vector $\thetavec$ has a standard complex Gaussian distribution, $\thetavec\sim\mathcal{CN}(\zerovec,\Imat)$. 
\item The analog, high-resolution measurements satisfy
\be \label{II.A.1}
	\xvec_a = \Hmat\thetavec + \uvec_a,
\ee
where $\Hmat = \left[\Hmat_1^T , \Hmat_2^T , \dots , \Hmat_{n_a}^T\right]^T \in \mathbb{C}^{N_a\times M}$, in which 
\be
\label{HH}
    \Hmat_n^H \Hmat_n = \rho_a \Imat_M, ~~~\forall n=1,\ldots,n_a,
\ee
$N_a = n_a M$, $\uvec_a \sim \mathcal{CN}(\zerovec,\sigma_a^2 \Imat_{N_a})$,  and $\sigma_a> 0$,  $\rho_a > 0$ are known.
If $i \neq j$, then $\Hmat_i^H \Hmat_j$  take arbitrary values. Thus, $\xvec_a|\thetavec \sim \mathcal{CN}(\Hmat\thetavec,\sigma_a^2 \Imat_{N_a})$.
\item The quantized, low-resolution measurements with a general threshold vector $\btau\in{\mathbb{C}}^{N_q}$ satisfy
\be	\label{II.A.2}
	\xvec_q = \mathcal{Q}\left(\Tmat\thetavec + \uvec_q -\btau\right),
\ee
where $\Tmat = \onevec_{n_q} \otimes \Tmat_1 \in \mathcal{C}^{N_q\times M}$, in which 
$N_q = n_q M$,
$\otimes$ is the Kronecker product, 
\be \label{III.A.4}
    \Tmat_1^H \Tmat_1 = \rho_q \Imat_M,
\ee
$\uvec_q \sim \mathcal{CN}(\zerovec,\sigma_q^2 \Imat_{N_q})$,
and $\sigma_q> 0$,  $\rho_q > 0$ are known.
The operator $\mathcal{Q}(\cdot)$ is defined in \eqref{I.1}. 
Under this model, 
\beqna \label{prob_q}
p(\xvec_q|\thetavec) \hspace{6.5cm}
\nonumber\\= \prod\limits_{n=1}^{N_q} 
\Phi (\zeta_n^R )^{\left( \frac{1}{2} + \frac{\Ree\{\xvec_{q_n}\}}{\sqrt{2}} \right)} 
\left( 1 - \Phi ( \zeta_n^R) \right)^{\left( \frac{1}{2} - \frac{\Ree\{\xvec_{q_n}\}}{\sqrt{2}} \right)}  \nonumber\\ \times
\Phi (\zeta_n^I )^{\left( \frac{1}{2} + \frac{\Imm\{\xvec_{q_n}\}}{\sqrt{2}} \right)} 
\left( 1 - \Phi (\zeta_n^I) \right)^{\left( \frac{1}{2} - \frac{\Imm\{\xvec_{q_n}\}}{\sqrt{2}} \right)},
\eeqna
where \(\phi(\cdot)\) and \(\Phi(\cdot)\) denote the standard Gaussian \ac{pdf} and \ac{cdf}, respectively. 
\item The noise  vectors,
 $\uvec_a$ and $\uvec_q$, and the parameters in $\thetavec$ are assumed to be mutually independent.
 \end{itemize}

The details on numerically computing the \ac{lmmse} and \ac{mmse} estimators under this model can be found in \cite{Mazor24}. 
In particular, by
substituting the prior and data probability functions \eqref{prob_q} in Algorithm \ref{Algo:MMSE} we obtain the \ac{mmse} estimator. In addition, it should be noted that 
for the case where $\bf{\tau} = \zerovec$, the \ac{lmmse} estimator has a closed-form expression \cite{berman2020resource}.

\subsection{WBCRB}
\label{WBCRB_LGO}
We demonstrated in \eqref{J_def} that for the specified model, the \ac{bfim} can be decomposed into prior, analog, and quantized contributions. Under the \ac{lgo} model, this decomposition takes the form given in \cite[Eq. (17)]{Mazor24}:
\be \label{fim_LGO}
 \Jmat =  
\Imat_M + \frac{1}{\sigma_a^2}\Hmat^H\Hmat + \frac{1}{2\sigma_q^2}\Tmat^H\E_{\thetavecsmall}[\Dmat(\thetavec)]\Tmat,  
\ee  
where  $\Tmat$ is defined before \eqref{III.A.4} and $\Dmat(\thetavec)  = \text{diag}\left([d_1(\thetavec),\cdots,d_{N_q} (\thetavec)]\right)\in \mathbb{R}^{N_q\times N_q}$, in which  
\be \label{dn}
d_n(\thetavec)\define   \frac{\phi^2(\zeta_n^R)}{\Phi(\zeta_n^R) \Phi(-\zeta_n^R)} + \frac{\phi^2(\zeta_n^I)}{ \Phi(\zeta_n^I) \Phi(-\zeta_n^I)},
\ee
\be
\label{zetaRI}
\zeta_n^R\define \frac{\sqrt{2}}{\sigma_q} {\text{\normalfont{Re}}}\{\tvec_n^T \thetavec - \tau_n\},~~~
\zeta_n^I\define \frac{\sqrt{2}}{\sigma_q} {\text{\normalfont{Im}}}\{\tvec_n^T \thetavec - \tau_n\},
\ee
and \(\tvec_n \in \mathbb{C}^{M \times 1}\) is the $n$th row of \(\Tmat\).

    We next examine the special cases from Subsection \ref{special_cases_subsection}
    
\subsubsection{BCRB}
The \ac{bcrb} from Subsection~\ref{special_bcrb} for this case is given by the inverse of \eqref{fim_LGO}, which also has a closed-form solution without requiring matrix inversion~\cite{Mazor24}.
As shown in \cite[p. 7]{van_bell}, the \ac{bcrb} is asymptotically tight {\em{iff}} the posterior \ac{bfim}, $\Jmat_{\xvec|\thetavecsmall}$, is independent of $\thetavec$. However, from \eqref{fim_LGO} it can be seen that the quantized component of the \ac{bfim}, $\Jmat_{\xvec_q|\thetavecsmall}$, depends on $\thetavec$. Consequently, the \ac{bcrb} is not asymptotically tight in this case, which highlights the need for an alternative bound.
Moreover, in \cite{Mazor24} we showed the limitation of the \ac{bcrb} for the \ac{lgo} model, which does not capture the non-monotonic behavior of the \ac{mse} \ac{wrt} \ac{snr}  in this case.

\subsubsection{\ac{fim}–Inverse Weighting}
\label{nice_wbcrb_for_lgo}
For this bound from Subsection \ref{case_at},
similar to the derivation of \eqref{fim_LGO}, it can be shown that \eqref{w_choice} under the \ac{lgo} model is reduced to 
\be \label{weight}
\Wmat(\thetavec)=\left(\thetavec\thetavec^H+\frac{1}{\sigma_a^2}\Hmat^H\Hmat + \frac{1}{2\sigma_q^2}\Tmat^H\Dmat(\thetavec)\Tmat\right)^{-1}.
\ee
To compute $\Gmat$ in this case, we  compute the derivatives of \eqref{weight} by using known derivative rules  (see, e.g., \cite[eq. 3.40]{complex_matrix}) as follows:
\beqna
\label{partial_w}
\frac{\partial [\Wmat(\thetavec)]_{k,m}}{\partial\theta_m^*}= 
-\left[\Wmat(\thetavec) \right.
\hspace{4cm}
\nonumber\\
\left. \times
\left( \thetavec \evec_m^T + \frac{n_q}{2\sigma_q^2} \sum_{n=1}^{M} \frac{\partial (d_n(\thetavec))}{\partial \theta_m^*}  \tvec_n^* \tvec_n^T \right)\Wmat(\thetavec)^H\right]_{k,m},
\eeqna
$k,m=1,\ldots, M$, for $\Wmat(\thetavec)$ from \eqref{weight}, where  $\evec_m \in \mathcal{R}^{M}$ is the $m$th column of the $M\times M$ identity matrix. 
In particular, for the scalar case (\(M=1\)), \eqref{partial_w} is reduced to 
\be \label{deri}
\frac{\ud w(\theta)}{\ud\theta^*}=-w(\theta)^2\left(\theta+\frac{n_q\rho_q}{2\sigma_q^ 2}\frac{\ud (d(\theta))}{\ud\theta^*}\right),
\ee
where we used \eqref{III.A.4}.
The regularity conditions from Section \ref{the_bound} are satisfied in this case too.

It should be noted that, similar to the \ac{bcrb}, obtaining a closed-form expression for the \ac{wbcrb} requires numerically evaluating the expectation (\ac{wrt} $\thetavec$) of the diagonal matrix $\Dmat(\thetavec)$, as well as the derivatives $\frac{\partial d_n(\thetavecsmall)}{\partial \theta_m^*}$. This evaluation can be carried out using numerical integration.
Nevertheless, the computation of this bound is significantly less demanding than computing the \ac{mmse} estimator from Subsection \ref{mmse_subsection}, as it avoids two-dimensional integration and does not require integration over the likelihood function or generating observations $[\xvec_a, \xvec_q]$. 
This reduces the complexity from high-dimensional integration to a tractable \(\mathcal{O}(K)\) procedure, where $K$ is the number of Monte Carlo simulations, making it feasible even for large $N_q$. 
In the supplemental material attached to this paper, we show that under the \ac{lgo} model there is no need to develop an augmented version of this bound.

Recalling that an asymptotically tight bound occurs {\em iff} $\Jmat_{\xvec|\thetavecsmall}$ in \eqref{FIMdata} is independent of $\thetavec$ (see Section~\ref{case_at}), and applying \eqref{weight} while neglecting the prior \ac{fim}, we obtain 
\be
\label{posterior_FIM_LGO}
\Jmat_{\xvec|\thetavecsmall}(\thetavec)=\frac{1}{\sigma_a^2}\Hmat^H\Hmat + \frac{1}{2\sigma_q^2}\Tmat^H\Dmat(\thetavec)\Tmat.
\ee
Since $\Jmat_{\xvec|\thetavecsmall}$ depends on $\thetavec$ in the presence of 1-bit data, the \ac{bcrb} is not asymptotically tight here, which explains the motivation for a tighter bound, such as the proposed \ac{wbcrb}.


\subsection{MSE Approximation}
\label{MSE_approx_LGO}

In this subsection, we derive a closed-form expression for the conditional probability of the event $\mathcal{N}$ (in which \eqref{eq:asymptotic_pmf} holds) under the \ac{lgo} model. This enables replacing the empirical approximation used in Algorithm~\ref{Algo:MC_region} with a semi-analytical computation.

Recall that $\mathcal{N}$ denotes the event in which all quantized measurements lie within a single cell. Under the \ac{lgo} model, and specifically, \eqref{II.A.2}, $\cvec(\thetavec)$ from \eqref{eq:decision_cell} is given by
\be
\cvec(\thetavec)={\mathcal{Q}}\left(\Tmat\thetavec -\btau\right).
\ee
Thus, 
\be
\Pr(\mathcal{N}) = \Pr\left(\mathcal{Q}(\Tmat\thetavec + \uvec_q -\btau)=\mathcal{Q}(\Tmat\thetavec -\btau)\right).
\ee

\noindent
Note that if $\btau$ is block-replicated as $\btau = \mathbf{1}_{n_q}\otimes \btau_1$, 
then $c(\thetavec) = \mathbf{1}_{n_q}\otimes \mathcal{Q}(\Tmat_1\thetavec - \btau_1)$.
Conditioned on $\thetavec$, consider each component $n$ of the quantized measurement.
Under the LGO model, the noise $u_{q,n}\sim\mathcal{CN}(0,\sigma_q^2)$ so that
$\Re\{u_{q,n}\},\Im\{u_{q,n}\}\stackrel{\mathrm{i.i.d.}}{\sim}\mathcal{N}(0,\sigma_q^2/2)$.
With the 1-bit quantizer in \eqref{I.1}, the real (resp. imaginary) output equals $+\tfrac{1}{\sqrt{2}}$
iff the noisy real (resp. imaginary) input is $\ge 0$, and $-\tfrac{1}{\sqrt{2}}$ otherwise.
Hence, the “no sign flip” events are
\[
{\text{sign}}\!\big(\Re\{\tvec_n^T\thetavec-\tau_n+u_{q,n}\}\big)
= {\text{sign}}\!\big(\Re\{\tvec_n^T\thetavec-\tau_n\}\big),
\]
\[
{\text{sign}}\!\big(\Im\{\tvec_n^T\thetavec-\tau_n+u_{q,n}\}\big)
= {\text{sign}}\!\big(\Im\{\tvec_n^T\thetavec-\tau_n\}\big).
\]
Using standard Gaussian tail relations, under the \ac{lgo} model, the corresponding conditional probabilities
(for the two outcomes per part) are
\begin{align}
\Pr\!\Big(\Re\{x_{q,n}\}=\pm\tfrac{1}{\sqrt{2}}\,\Big|\,\thetavec\Big)&=\Phi(\pm\zeta_n^R),\label{eq:real_probs}
\\
\Pr\!\Big(\Im\{x_{q,n}\}=\pm\tfrac{1}{\sqrt{2}}\,\Big|\,\thetavec\Big)&=\Phi(\pm\zeta_n^I),\label{eq:imag_probs}
\end{align}
where $\zeta_n^R,\zeta_n^I$ are defined in \eqref{zetaRI}.
Using conditional independence across $n$ and between real and imaginary parts given $\thetavec$, the conditional probability that all quantized measurements match their noiseless decision cell is
\begin{align}
\Pr(\mathcal{N}|\thetavec)
&=\prod_{n=1}^{N_q}\Pr\!\big(\text{no flip in real}|\thetavec\big)\;
                          \Pr\!\big(\text{no flip in imag}|\thetavec\big) \nonumber\\
&=\prod\nolimits_{n=1}^{N_q}\Phi\!\big(|\zeta_n^R|\big)\,\Phi\!\big(|\zeta_n^I|\big).
\label{eq:PrN_theta}
\end{align}
Finally, by substituting \eqref{eq:PrN_theta} in  \eqref{prob_n}, we obtain 
\begin{equation}
\Pr(\mathcal{N})
=\E_{\thetavecsmall}\!\left[\prod\nolimits_{n=1}^{N_q}\Phi\!\big(|\zeta_n^R|\big) \times \Phi\!\big(|\zeta_n^I|\big)\right],
\label{Q_sim}
\end{equation}
where \eqref{Q_sim} computed by averaging over the prior distribution of $\thetavec$. In practice, this is implemented via Monte Carlo simulations \ac{wrt} the prior of $\thetavec$,
and reduces the problem to sampling over $\thetavec$, rather than evaluating estimator performance. 
Importantly, in the high-\ac{snr} regime where $\sigma_q \to 0$, 
whenever $\Re\{\tvec_n^T\thetavec-\tau_n\}$ or $\Im\{\tvec_n^T\thetavec-\tau_n\}$ is nonzero,
we have $\Phi(|\zeta_n^R|),\Phi(|\zeta_n^I|)\to 1$, hence the overall probability $\Pr(\mathcal{N})$ converges to $1$; the measure-zero case of exactly zero mean does not alter the integral under a continuous prior.
This behavior formally captures the saturation effect of quantized measurements: at sufficiently high \ac{snr}, all quantized outputs collapse to a deterministic constant value, so that  $\Pr(\mathcal{N})\rightarrow 1$, 
and the quantized data become asymptotically uninformative compared to the analog data.


\textbf{Remark:} For a pure 1-bit system, $n_a=0$, it can be shown based on \eqref{eq:posterior_cell_1_bit}, that the \ac{mse} of the \ac{lmmse} and \ac{mmse} estimators converge to \cite{Benedikt23}
\be \label{1-bit_lmmse}
\lim_{\sigma_q \to 0}{\mathbf{MSE}}(\hat{\thetavec}_{\text{MMSE}}(\xvec_q)) = (1-\frac{2}{\pi})\Imat_M.
\ee
By substituting \eqref{Q_sim}, \eqref{1-bit_lmmse},  and the \ac{wbcrb} 
into \eqref{analog_app}, we  obtain the \ac{mse} approximation for this case.

\section{Simulations} \label{sim} 

In this section, we evaluate the proposed bounds and the MSE approximation method under the \ac{lgo} model described in Section~\ref{model_LGO}. We compare them with the \ac{mse} of the \ac{mmse} and \ac{lmmse} estimators. 
The \ac{mmse} estimator is implemented using Algorithm \ref{Algo:MMSE}. The \ac{lmmse} estimator is implemented using the closed-form expression from~\cite{berman2020resource} for $\tau = 0$,  and via Algorithm~1 in \cite{Mazor24} for general threshold values.
The simulations are performed for the scalar case $M = 1$, which enables feasible implementation of the \ac{mmse} estimator.

The following performance analysis tools are examined:
\begin{itemize}
  \item The \ac{bcrb} from  \eqref{bcrb}, which, under the \ac{lgo} model, is given by the inverse of \eqref{fim_LGO}, computed without matrix inversion as in~\cite{Mazor24}.
    \item The \ac{wbcrb} based on the \ac{fim}-inverse weighting  from Subsection \ref{case_at}, computed under the \ac{lgo} model using \eqref{weight}, denoted as ``WBCRB".
   \item The optimal \ac{wbcrb}, obtained from \eqref{opt_wbcrb} with $\Delta=0.2$ and $L=1,000$ test points, denoted as ``WBCRB opt".
\item  The proposed \ac{mse} approximation method \eqref{analog_app}, where the saturation regime probability is computed using \eqref{Q_sim}.
\end{itemize}

Unless stated otherwise, all results are obtained by $1000$ Monte Carlo trials, with parameters
$\sigma_a^2= \sigma_q^2= \sigma^2$, $\rho_a= \rho_q = 1$, $N_t=1,000$,
and $M=1$, where $\ac{snr} \triangleq \rho / \sigma$.

\subsection{Test Case 1: Effect of the Threshold}
\begin{figure}[hbt]
    \centering
	\subcaptionbox{\label{Fig3a}}[\linewidth]	{\includegraphics[width=1.05\linewidth]{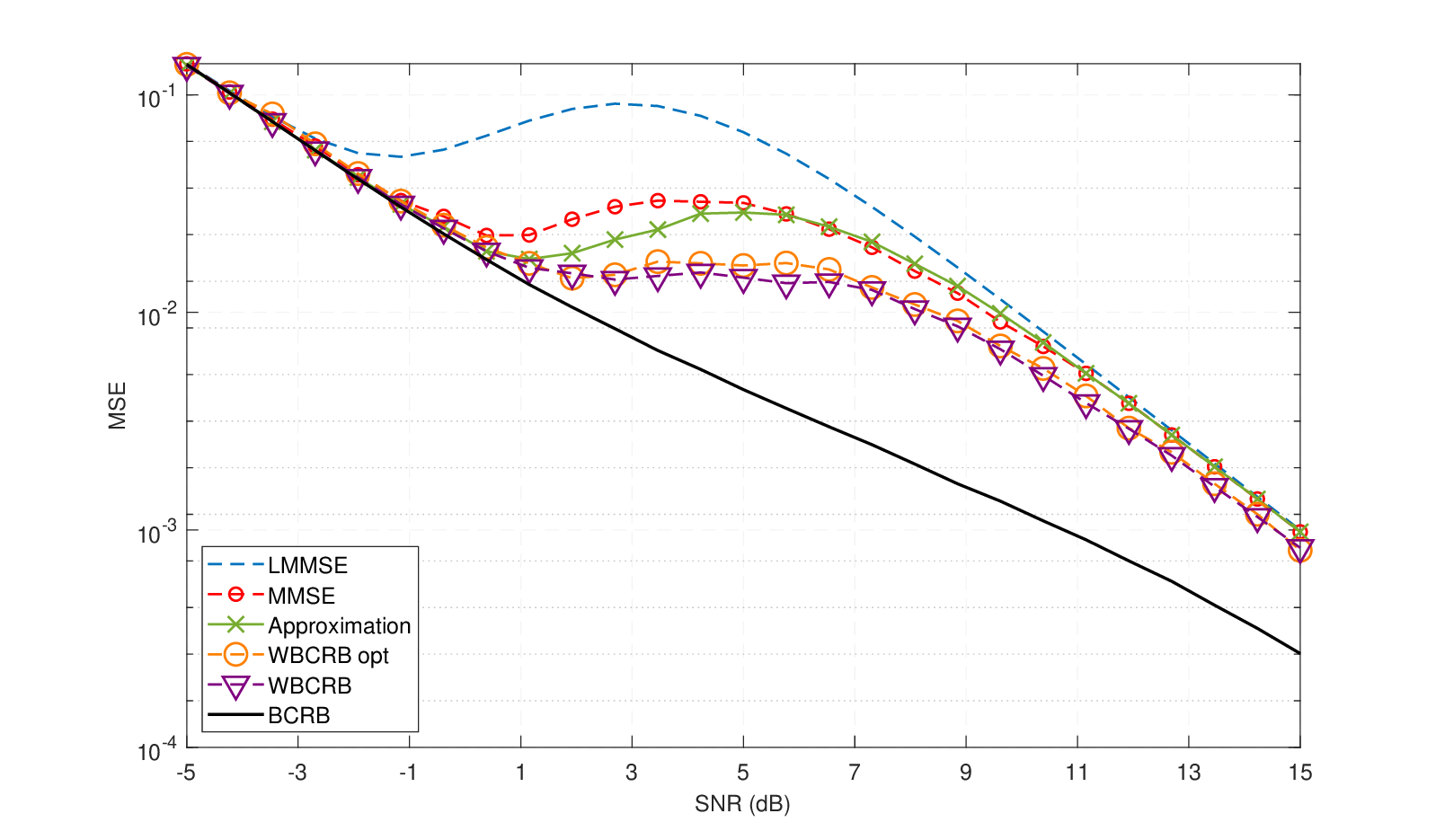}	}
        \subcaptionbox{\label{Fig3b}}[\linewidth]	{\includegraphics[width=1.05\linewidth]{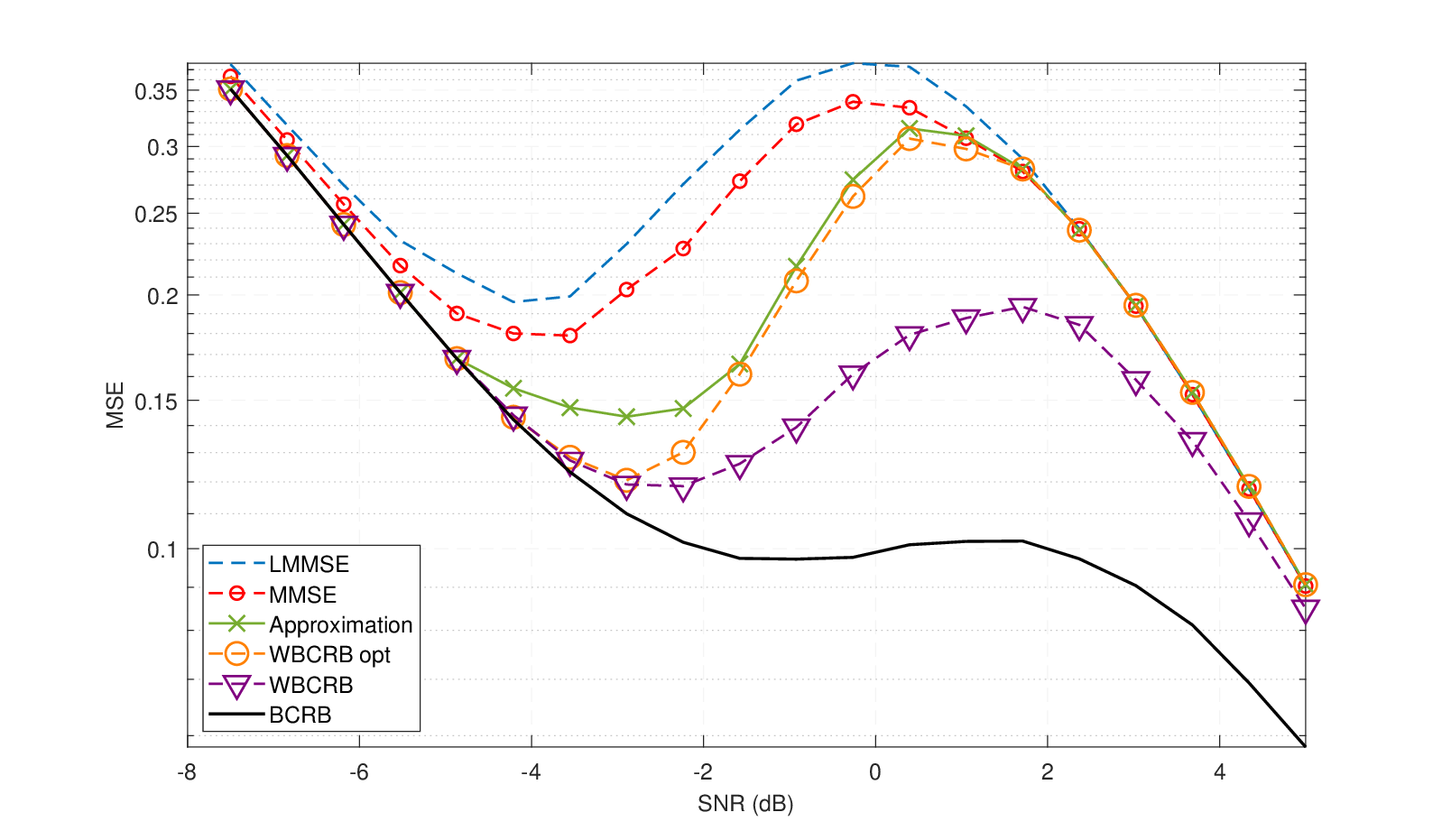}}
    \caption{\ac{mse} of the \ac{mmse} and \ac{lmmse} estimators in comparison to  different \ac{wbcrb} weight selection versus \ac{snr} for $(n_a,n_q)=(1,100)$ and (a) $\tau=0$ (b) $\tau=2$. }
	\label{Fig1}
\end{figure}

In Figs. \ref{Fig3a} and~\ref{Fig3b}, we present the \ac{mse} of the \ac{mmse} and \ac{lmmse} estimators,
along with the different bounds, versus \ac{snr} 
for $n_a=1$, $n_q=100$, and two threshold settings: $\tau=0$ (Fig. \ref{Fig3a}) and $\tau=2.5$ (Fig. \ref{Fig3b}). 
In both cases, the \ac{mse}  curves exhibit non-monotonic dependence on \ac{snr}, reflecting the varying informativeness of the quantized measurements. 
For $\tau = 0$ (Fig.~\ref{Fig3a}), the \ac{bcrb} decreases monotonically with \ac{snr} and fails to capture the region where the quantized data becomes non-informative.
For $\tau = 2.5$ (Fig.~\ref{Fig3b}), the \ac{bcrb} itself exhibits non-monotonic behavior, providing a better qualitative match to the estimator \ac{mse} trends. However, in this case the overall \ac{mse} is significantly higher, making such thresholds undesirable from a performance standpoint.
These results reveal a trade-off: the \ac{bcrb} is tighter when estimator performance is degraded, which emphasizes the need for tighter bounds. 
In addition, it can be seen that the FIM–inverse weighting \ac{wbcrb} and the optimal \ac{wbcrb} exhibit non-monotonic behavior and track the different regions of the \ac{mse} of the estimators with high accuracy. 
In Fig. \ref{Fig3a}, the FIM–inverse weighting \ac{wbcrb} is close to the optimal \ac{wbcrb}, which makes it a more practical choice in this setting. 
The proposed \ac{mse} approximation method further improves prediction accuracy for both cases, and closely matches the \ac{mmse} performance. 
At high \ac{snr}, all estimators and bounds converge, as the estimation relies solely on the analog observations under Gaussian noise.
Regarding the \ac{mse} approximation, when ${\boldsymbol\tau} \approx \zerovec$, the proposed method is tighter than the \ac{wbcrb}, and closely matches the \ac{mmse} estimator.

\begin{figure}[hbt]    \centering{\includegraphics[width=1.05\linewidth]{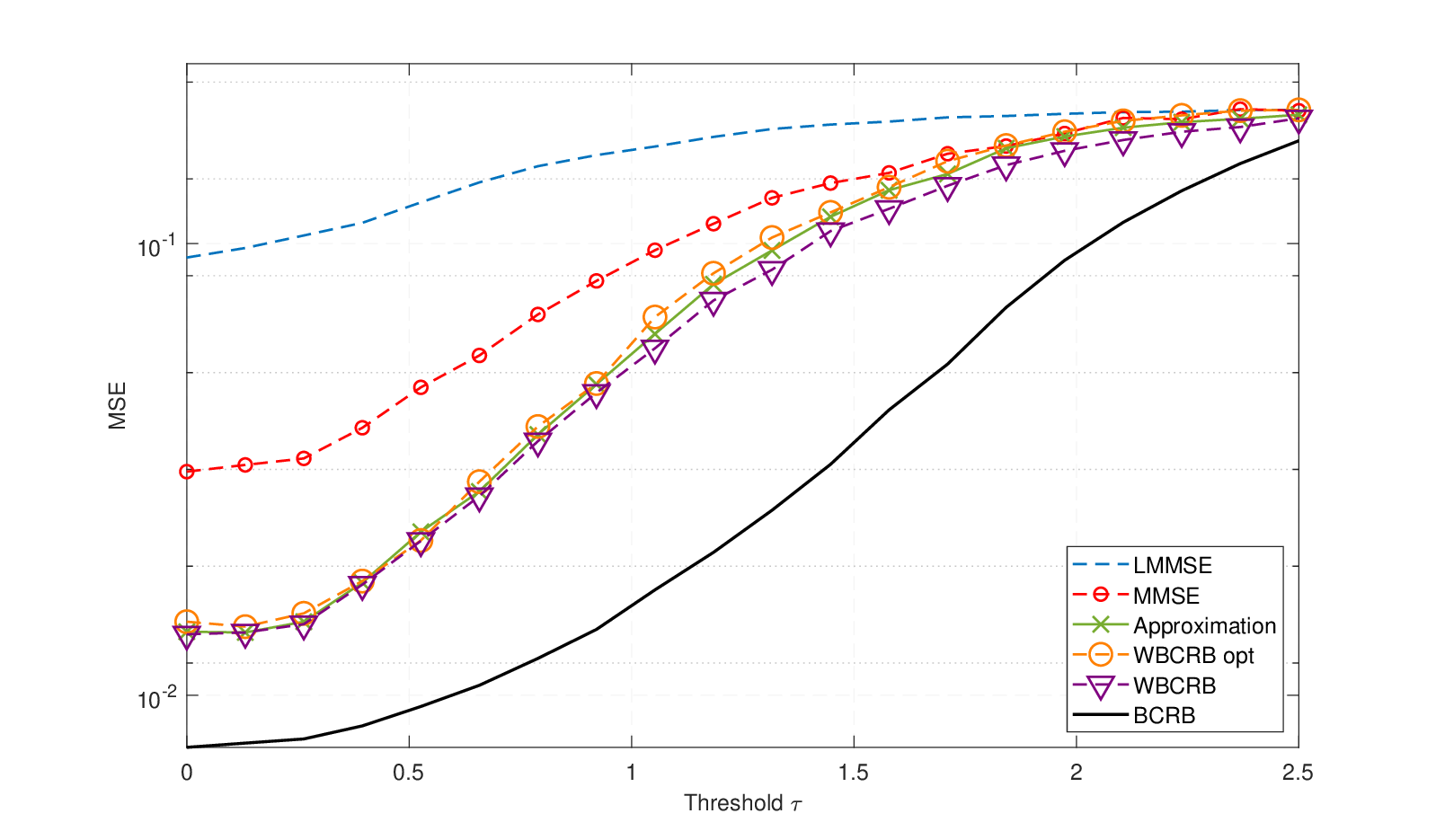}}
    \caption{\ac{mse} of the \ac{mmse} and \ac{lmmse} estimators in comparison to different \ac{wbcrb} weight selection versus $\tau$ for $(n_a,n_q)=(1,100)$ and $\sigma=\frac{1}{2}$}
	\label{fig_thresh}
\end{figure}

Figure~\ref{fig_thresh} complements this analysis by showing the \ac{mse} and bounds as a function of the quantization threshold $\tau$ for $\sigma = 1/2$ and $(n_a, n_q) = (1, 100)$. 
As $\tau$ increases, the system approaches the saturation region, in which most quantized outputs collapse to identical values, leading to saturation in both the \ac{mse} and the \ac{wbcrb} variants. In the mid-threshold region of $\tau \in [0.5, 2]$, the  FIM–inverse weighting \ac{wbcrb} is significantly tighter and follows the \ac{mmse} more closely than the conventional \ac{bcrb},  highlighting its advantage in accurately capturing estimation performance in transition regions between informative and non-informative quantized measurements.

\subsection{Test Case 2: 1-Bit System} \label{onebit_sim} 
\begin{figure}[hbt]
    \centering{\includegraphics[width=1.05\linewidth]{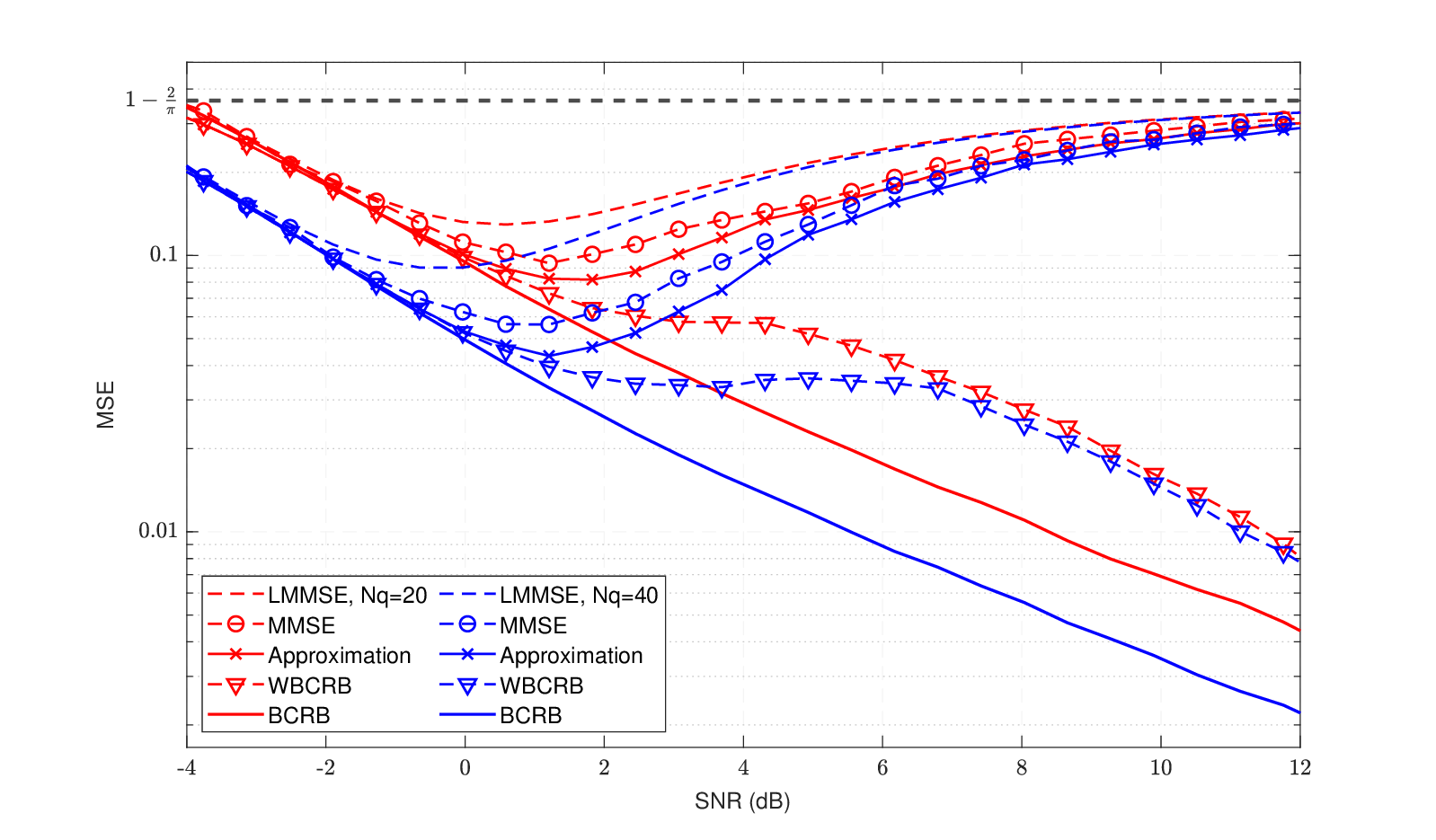}}
     \caption{\ac{mse} of the \ac{lmmse} and \ac{mmse} estimators in comparison with  \ac{wbcrb} and \ac{bcrb} versus \ac{snr} for different pure 1-bit partitions, $n_a=0$, and $\tau=0$.}
	\label{fig_onebit}
\end{figure}

In Fig.~\ref{fig_onebit}, we present the \ac{mse} of the \ac{lmmse} and \ac{mmse} estimators versus \ac{snr} for different values of $n_q$ (20 and 40), under a pure 1-bit system  ($n_a = 0$) and $\tau = 0$. 
The optimal \ac{wbcrb}  is omitted for clarity, as its behavior closely matches that of FIM–inverse weighting \ac{wbcrb} in this case. 
It can be seen that as the \ac{snr} increases, the performance of both estimators converges to a constant value, which is independent of $n_q$, consistent with the saturation phenomenon discussed in Sections~\ref{Approximaiton_Method} and~\ref{MSE_approx_LGO}. This behavior arises from the loss of information in quantized measurements at high \ac{snr} (cf. event $\mathcal{N}$ \eqref{eq:asymptotic_pmf}).
The \ac{bcrb} fails to reflect this trend, decreases monotonically with \ac{snr},
 and maintains an artificial gap between different sensor configurations ($n_q=20$ and $n_q=40$) even asymptotically.
In contrast, the \ac{wbcrb} captures the non-monotonic behavior and the convergence of performance across $n_q$ values in the high-\ac{snr} regime, although it is not a tight bound.
 Notably, the proposed \ac{mse} approximation almost perfectly matches the \ac{mmse} curve over the entire \ac{snr} range and converges to the theoretical limit $1-\frac{2}{\pi}$ at high \ac{snr} (see \eqref{1-bit_lmmse}). This figure thus highlights the importance of the proposed approximation method as a reliable and computationally efficient tool for accurately predicting estimation performance when quantization is involved.

\subsection{Resource Allocation} \label{resource_subsec}
\begin{figure}[hbt]
    \centering{\includegraphics[width=1.05\linewidth]{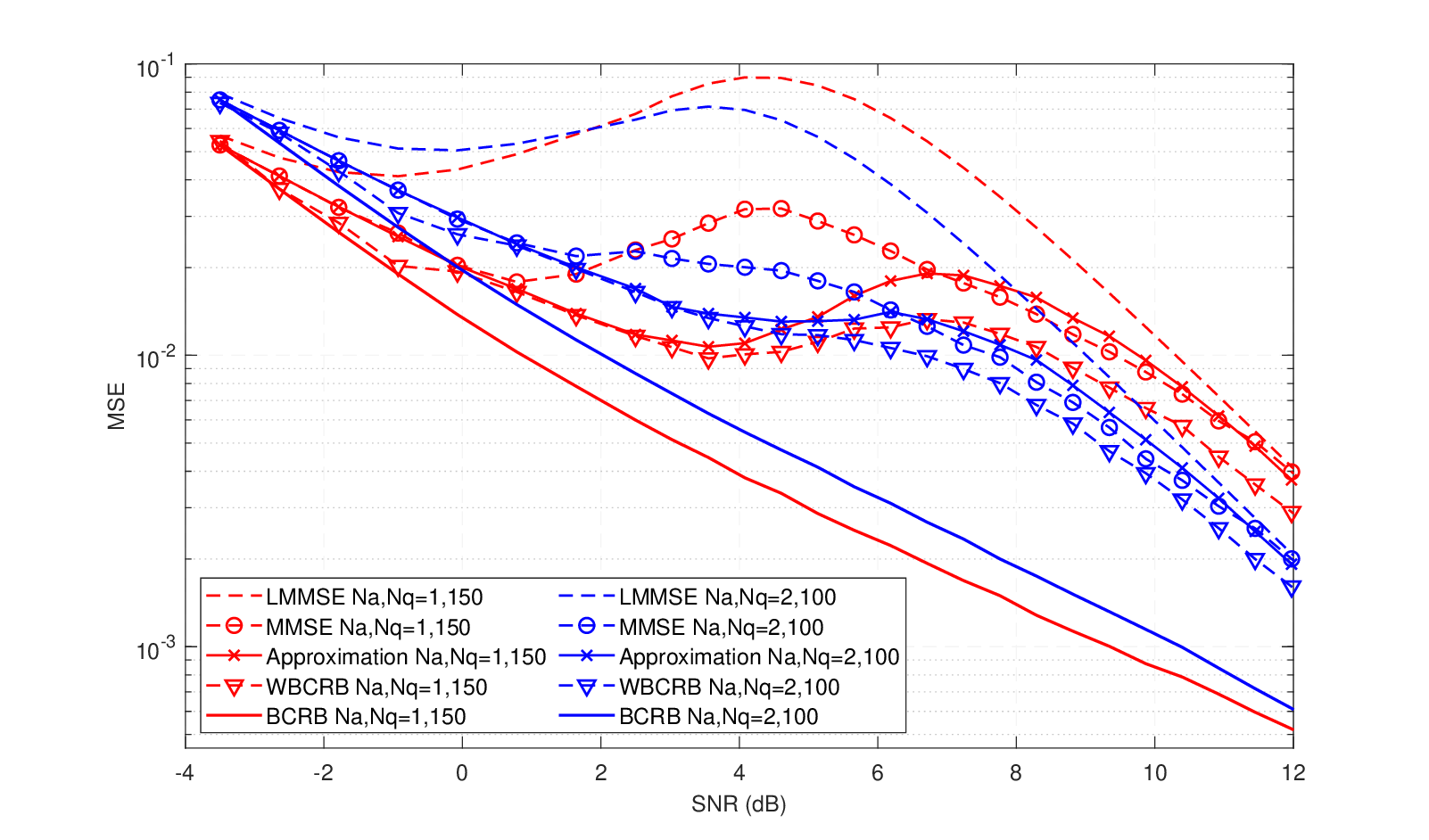}}
    
    \caption{The \ac{mse} of the  \ac{bcrb} and \ac{mmse} estimator in comparison with the proposed MSE approximation as a function of the SNR for different numbers of measurements.}
	\label{resource2}
\end{figure}

We next examine the use of the proposed \ac{mse} approximation \eqref{MSE_approx_LGO} as a practical design tool for resource allocation between analog and quantized measurements in mixed-resolution systems. Resource allocation plays a central role in determining the performance–power trade-off, as the partition $(n_a,n_q)$ directly impacts estimation accuracy across different \ac{snr} regimes.
Figure~\ref{resource2} shows the \ac{mse} of the estimators, the proposed \ac{mse} approximation, and the mixed-resolution \ac{bcrb} for two measurement partitions: $(n_a,n_q) = (1,150)$ and $(n_a,n_q) = (2,100)$. While the \ac{mmse} curves reveal that the optimal partition depends on the \ac{snr}, with an intersection at $\ac{snr} \approx 2.5$, and analog-heavy configurations outperforming at high \ac{snr}, the \ac{bcrb} fails to capture these trends. In particular, at high \ac{snr} the \ac{bcrb} reveals no change in relative performance, even though quantized data becomes non-informative and the number of analog measurements dominates accuracy.
In contrast, the \ac{wbcrb}, used as the basis for the proposed approximation, captures the impact of the quantized data, enabling a more accurate prediction of performance trends than the \ac{bcrb}. 
Moreover,  the proposed \ac{mse} approximation, which builds on the \ac{wbcrb} from \eqref{weight}, closely tracks the \ac{mmse} across all \ac{snr} regions and correctly identifies the more effective partition in each region, with intersection at $\ac{snr} \approx 5$ (as predicted by the \ac{wbcrb}). This figure thus highlights the proposed approximation as a reliable tool for mixed-resolution system design.


\section{Conclusions} \label{conc}
In this paper, we developed the \ac{wbcrb} for mixed-resolution Bayesian estimation using both quantized and analog data. Several weighting strategies were investigated, including the conventional \ac{bcrb}, a data-aware \ac{wbcrb} based on the inverse instantaneous \ac{bfim}, and the optimal \ac{wbcrb}~\cite{Joseph24}.  
We also proposed an SNR-dependent \ac{mse} approximation method that partitions the parameter space according to the informativeness of the quantized measurements, expressing the overall \ac{mse} as a weighted combination of region-specific \acp{wbcrb} and an analog-only estimator. As an illustrative case, the \acp{wbcrb} and MSE approximation were derived for the widely used \ac{lgo} model.
Simulation results showed that the proposed \ac{wbcrb} variants closely track the \ac{mse} of the \ac{mmse} estimator, substantially outperforming the \ac{bcrb}. The improvement is most pronounced in the mid-SNR region, where quantization effects lead to non-monotonic \ac{mse} behavior: while the \ac{bcrb} fails to capture these transitions, the two \acp{wbcrb} reliably predict them. The SNR-dependent \ac{mse} approximation further demonstrates even higher accuracy compared to the bounds, offering a practical performance prediction tool.
Overall, the results highlight the value of the \ac{wbcrb} framework for accurate performance analysis and resource allocation in mixed-resolution systems.

\vspace{-0.75cm}
\appendices

\section{Proof of Theorem \ref{theo_1}}
\label{AppA}
The proof is along the lines of the proof in \cite{Joseph24}, adapted for the mixed-resolution setting, e.g., it explicitly incorporates summations for the quantized data. For simplicity, in this appendix we denote an arbitrary estimator of $\thetavec$ by $\hat{\thetavec}(\xvec_a, \xvec_q)$.

\textbf{Step 1 - Definition of the Auxiliary Function:}
Let the vector function
$\gvec(\xvec_a,\xvec_q,\thetavec) \in \mathbb{C}^M$
be
\beqna
\label{g_def}
\gvec(\xvec_a,\xvec_q,\thetavec) = \Wmat(\thetavec)\nabla^H_{\thetavecsmall}\left\{\log p(\thetavec)p(\xvec_a|\thetavec)p(\xvec_q|\thetavec)\right\} + \vvec(\thetavec),
\eeqna
where the elements of $\vvec(\thetavec)$ are defined in \eqref{v_def}.
According to Condition \ref{cond1} and the differentiability of $\Wmat(\thetavec)$, $\gvec(\xvec_a,\xvec_q,\thetavec)$ is well-defined and is differentiable \ac{wrt} $\thetavec$.
Next, we verify that it satisfies the crucial unbiasedness property \cite{bobrovsky87,van_bell}: 
\be
\label{g_zero}
\E_{\thetavecsmall|\xvec_a,\xvec_q}[\gvec(\xvec_a,\xvec_q,\thetavec)\big| \xvec_a, \xvec_q] = \zerovec.
\ee 
For each $k$, we have
\beqna \nonumber
\E_{\thetavecsmall|\xvec_a,\xvec_q}[g_k^*(\xvec_a,\xvec_q,\thetavec)|\xvec_a,\xvec_q] %
\hspace{3cm}
\\=\int_{\Omega_\thetavecsmall}\sum_{m=1}^M
\left([\Wmat^*(\thetavec)]_{k,m} \frac{\partial \log\left(p(\cdot)\right)}{\partial \theta_m} \right.
\hspace{1cm}\nonumber\\
 \label{mixedg}
 \left.
+\frac{\partial[\Wmat^*(\thetavec)]_{k,m}}{\partial\theta_m}\right)  p(\thetavec|\xvec_a,\xvec_q)d\thetavec. 
\eeqna 
 Using the 
 derivative of the log and interchanging the order of summing and integration (see Condition~\ref{cond2}), one obtains
 \beqna 
\E_{\thetavecsmall|\xvec_a,\xvec_q}\left[g_k^*(\xvec_a,\xvec_q,\thetavec)|\xvec_a,\xvec_q\right]
\hspace{2.75cm}\nonumber
\\ 
=\sum_{m=1}^M \int_{\Omega_\thetavecsmall} \frac{\partial [\Wmat^*(\thetavec)]_{k,m}}{\partial \theta_m} p(\thetavec|\xvec_a,\xvec_q) d\thetavec 
\hspace{0.5cm}\nonumber\nonumber\\ 
+ \sum_{m=1}^M \int_{\Omega_\thetavecsmall} [\Wmat^*(\thetavec)]_{k,m} \frac{\partial p(\thetavec|\xvec_a,\xvec_q)}{\partial \theta_m} d\thetavec \hspace{0.25cm}
\nonumber\\ 
=\sum_{m=1}^M \int_{\Omega_\thetavecsmall}\frac{\partial}{\partial\theta_m}\big([\Wmat^*(\thetavec)]_{k,m}p(\thetavec|\xvec_a,\xvec_q)\big).
\eeqna
Applying integration by parts to the last term, and using \eqref{con_a}, the boundary terms vanish and
we get \eqref{g_zero}.
Thus, the vector-valued function $\gvec(\xvec_a,\xvec_q,\thetavec)$ is properly constructed for use in the Cauchy-Schwarz-based bound derivation in the following.


{\bf{Step 2 - Cauchy-Schwarz inequality and derivation of \eqref{Gmat}:}}
Let $\hat{\thetavec}$ be any estimator of $\thetavec$. Using the Cauchy-Schwarz inequality \cite[Eq.\,(189)]{van_bell}, the \ac{mse} matrix satisfies
\beqna \label{inq_final}
\E \left[ (\hat{\thetavec}- \thetavec)( \hat{\thetavec}- \thetavec)^H\right] 
\succeq \E\left[ ( \hat{\thetavec}- \thetavec) \gvec^H(\xvec_a,\xvec_q,\thetavec) \right] \hspace{1cm}
\nonumber\\ \times \tilde{\Gmat}^{-1} \E\left[  \gvec(\xvec_a,\xvec_q,\thetavec) ( \hat{\thetavec}- \thetavec)^H
\right],
\eeqna
where
\begin{equation} \label{g_mat}
    \tilde{\Gmat} = \E \left[ \gvec(\xvec_a,\xvec_q,\thetavec) \gvec^H (\xvec_a,\xvec_q,\thetavec)  \right].
\end{equation}
It should be noted that we use the fact that, according to \eqref{cond2}
 $\mathbf{g}(\xvec,\thetavec) $  in \eqref{g_def}  is a measurable function that satisfies 
\eqref{g_zero}, and thus, the use of \eqref{inq_final} is valid \cite{van_bell}.


Substituting  $\gvec(\xvec_a, \xvec_q, \thetavec)$   from \eqref{g_def} 
in \eqref{Gmat}, results in 
\begin{align}
\label{GGG}
\tilde{\Gmat} 
&= \E\big[ \Wmat(\thetavec) \nabla_{\thetavecsmall}^H \log p(\cdot) \nabla_{\thetavecsmall} \log p(\cdot) \Wmat^H(\thetavec) \big] \nonumber\\
&\quad + \E\big[ \Wmat(\thetavec) \nabla_{\thetavecsmall}^H \log p(\cdot)\vvec^H(\thetavec) \big] \nonumber\\
&\quad + \E\big[ \vvec(\thetavec) \nabla_{\thetavecsmall} \log p(\cdot) \Wmat^H(\thetavec) \big] \nonumber\\
&\quad + \E\big[ \vvec(\thetavec)\vvec^H(\thetavec) \big],
\end{align}
where in this appendix $p(\cdot)$ is used as a shorthand for $p(\thetavec)p(\xvec_a|\thetavec)p(\xvec_q|\thetavec)$.
Using Condition~\ref{cond4},
the $(n,k)$th element of the third term 
in \eqref{GGG}
can be written as
\beqna \label{a_mat}
\E \left[  v_n(\thetavec)[\nabla_{\thetavecsmall}\log p(\thetavec) \Wmat^H(\thetavec)]_k \right]\hspace{2.5cm}
\nonumber\\
=\E \left[ \sum_{m=1}^M
\Wmat^*(\thetavec)_{k,m}\frac{\partial \log p(\thetavec)}{\partial \theta_m} v_n(\thetavec)  \right]
\nonumber\\
=
\sum_{m=1}^M\int_{\Omega_\thetavecsmall} \frac{\partial p(\thetavec)}{\partial \theta_m}[\Wmat(\thetavec)]^*_{k,m} v_n (\thetavec) d \thetavec,
\eeqna
where the sum and integration may be interchanged under Condition~\ref{cond2}. 
Using integration by parts for the inner expression and using the condition in \eqref{con_b}, we obtain that each marginal integral in \eqref{a_mat} satisfies
\beqna \label{parts}\int_{[\Omega_{\thetavecsmall}]_m}\frac{ \partial p(\thetavec)}{\partial \theta_m}[\Wmat(\thetavec)]^*_{k,m} v_n(\thetavec)\, d\theta_m
\hspace{2cm}\nonumber\\
=
-\int_{[\Omega_{\thetavecsmall}]_m} \frac{\partial \left([\Wmat(\thetavec)]^*_{k,m}v_n(\thetavec)\right)}{\partial \theta_m} p(\thetavec) \, d\theta_m.
\eeqna
Substitution of \eqref{parts} into \eqref{a_mat} results in
\beqna \label{G_proof}
\E \left[  v_n(\thetavec)[\nabla_{\thetavecsmall}\log p(\thetavec) \Wmat^H(\thetavec)]_k \right]
=[\Amat]^*_{k,n},
\eeqna
where the elements of $\Amat$ are defined in \eqref{A_def}. Similarly, the second term in \eqref{GGG} equals $\Amat$.
Substituting these results in \eqref{GGG}, we obtain $\tilde{\Gmat}=\Gmat$, where $\Gmat$ is defined in \eqref{Gmat}.
Note that this derivation of $\Amat$ leads to an expression for $\Gmat$ that depends solely on the weight matrix $\Wmat(\thetavec)$.

\textbf{Step 3 - Evaluation of the Mixed Term:}
To complete the proof, we need to compute the cross-moment in \eqref{inq_final}.
First, using \eqref{g_def} one obtains
\beqna
\label{mixed_term}
\E\left[(\hat{\thetavec}-\thetavec)\,\gvec^H(\xvec_a,\xvec_q,\thetavec)\right] = 
\int\limits_{\Omega_\thetavecsmall} \hspace{-0.1cm}\sum_{\xvec_q \in \mathcal{Z}^{N_q}}\int_{\mathbb{C}^{N_a}} (\hat{\thetavec}-\thetavec)\, \nonumber\\
 \times \left(\Wmat(\thetavec) \nabla^H_{\thetavecsmall} \log p(\cdot) +\vvec(\thetavec)\right) p(\cdot) d\xvec_a d\thetavec,
\eeqna
where  the interchange of sum and integral is justified by Condition~\ref{cond2}.
The $(n,k)$th element of this matrix is given by
\beqna 
\label{38_eq}
\E\left[(\hat{\theta}_n-\theta_n)\,g_k^*(\xvec_a, \xvec_q, \thetavec)\right] 
\hspace{3.5cm}\nonumber\\=
 \sum_{\xvec_q \in \mathcal{Z}^{N_q}} \int_{\mathbb{C}^{N_a}} \sum_{m=1}^M \int_{\Omega_{\thetavecsmall}}(\hat{\theta}_n-\theta_n) 
\hspace{2.5cm}\nonumber\\\times
 \big[[\Wmat^*(\thetavec)]_{k,m} \frac{\partial \log\left(p(\cdot)\right)}{\partial \theta_m} 
+\frac{\partial[\Wmat^*(\thetavec)]_{k,m}}{\partial\theta_m}\big] p(\cdot)
\, d\thetavec d\xvec_a
\nonumber\\
= \sum_{\xvec_q \in \mathcal{Z}^{N_q}} \int_{\mathbb{C}^{N_a}} \sum_{m=1}^M \int_{\Omega_{\thetavecsmall}}(\hat{\theta}_n-\theta_n) 
\hspace{2cm}\nonumber\\ \times
 \big[[\Wmat^*(\thetavec)_{k,m}] 
 \frac{\partial p(\cdot)}{\partial \theta_m}+p(\cdot)\frac{\partial {[\Wmat^*(\thetavec)_{k,m}] }}{\partial \theta_m}\big] 
\, d\thetavec d\xvec_a.
\eeqna
Integration by parts over $\theta_m$ on the r.h.s. in \eqref{38_eq} yields
\be \label{T_inner}
-\int\delta_{n,m}[\Wmat^*(\thetavec)]_{k,m}p(\thetavec)p(\xvec_a|\thetavec)p(\xvec_q|\thetavec) d\theta_m,
\ee
where $\delta_{n,m}$ is the Kronecker delta,
and due to \eqref{con_a}, the boundary terms vanish.
It should be noted that we used
 the fact that for any differentiable function $f(\thetavec)$,
\[
\theta_n \frac{\partial f(\thetavec)}{\partial \theta_m} = \frac{\partial}{\partial \theta_m} \left\{ \theta_n f(\thetavec) \right\} - \delta_{n,m} f(\thetavec).
\]
 By substituting \eqref{T_inner} in \eqref{mixed_term}, and using the fact that $\Wmat(\thetavec)$ is a Hermitian matrix, we obtain
\be
\label{T_proof}
\E[(\hat{\thetavec} - \thetavec) \gvec^H(\xvec_a, \xvec_q, \thetavec)] = -\E[\Wmat^H(\thetavec)].
\ee
By substituting $\tilde{\Gmat}=\Gmat$ (see after \eqref{G_proof}),  \eqref{G_proof}, and \eqref{T_proof} in \eqref{inq_final}, we obtain the \ac{wbcrb} in \eqref{bound_wbcrb}.


{\Large{Supplemental Material for  ``Weighted Bayesian Cram$\acute{\text{e}}$r–Rao Bound for Mixed-Resolution Parameter Estimation"}}

\vspace{1cm}

 This document contains supplemental material for the paper. 
In the following, we present an extension of the proposed \ac{wbcrb} in \eqref{weight}, and show that under the \ac{lgo} model, it is sufficient to use only the standard WBCRB.

\section{Extension of the WBCRB on the Augmented MSE} \label{complex}
In the estimation of complex-valued parameters, the conventional \ac{mse} matrix is insufficient to fully characterize second-order statistics \cite{Ollila08}. 
 Instead, both the covariance and pseudo-covariance matrices should be considered.
 To this end, we define the augmented form as 
 \[\bar{\epsilonvec} = \left[(\hat{\thetavec}(\xvec_a, \xvec_q) - \thetavec)^T,(\hat{\thetavec}(\xvec_a, \xvec_q) - \thetavec)^H 
\right]^T.
 \]
The augmented \ac{mse} matrix
is defined as
\begin{equation}
\label{eq:augmented_mse}
\mathbf{MSE}_{\text{aug}} = \E\left[ (\bar{\epsilonvec} - \E[\bar{\epsilonvec}])(\bar{\epsilonvec} - \E[\bar{\epsilonvec}])^H \right] \in \mathbb{C}^{2M \times 2M}.
\end{equation}
It can be seen that the upper-left $M\times M$ block of $\mathrm{MSE}_{\text{aug}}$ is the MSE matrix,
$\E \left[ (\hat{\thetavec}(\xvec_a, \xvec_q) - \thetavec)(\hat{\thetavec}(\xvec_a, \xvec_q) - \thetavec)^H \right]$, which is bounded in Theorem \ref{theo_1}.
The lower-right $M\times M$ block is the complex-conjugate of the \ac{mse} matrix. The off-diagonal $M\times M$ blocks are the pseudo-covariance of the estimation error vector, given by
 $\E \left[ (\hat{\thetavec}(\xvec_a, \xvec_q) - \thetavec)(\hat{\thetavec}(\xvec_a, \xvec_q) - \thetavec)^T \right]$, and its complex conjugate. It should be noted that even if the true parameter vector $\thetavec$ is circularly symmetric (i.e., proper), the estimation error may still be an improper complex random vector, since the estimator is not constrained to be proper. In such cases, the conventional covariance matrix does not fully capture the second-order statistics, and it is necessary to consider the augmented error and provide a lower bound on the full augmented MSE matrix.

We now present the WBCRB on the augmented MSE~\eqref{eq:augmented_mse} in a similar manner to Theorem~\ref{theo_1}, but using the augmented \ac{bfim}.  
In this case, we use the Hermitian, positive-definite weight matrix 
$\bar\Wmat(\thetavec)\in\mathbb{C}^{2M\times2M}$. Similar to \eqref{v_def}, we define
the vector $\bar\vvec(\thetavec)\;\in\;\mathbb{C}^{2M}$ with components
\[
[\bar\vvec(\thetavec)]_k
=\sum_{m=1}^{2M}\frac{\partial[\bar\Wmat(\thetavec)]_{k,m}}{\partial\bar\theta_m^*},
\]
where
$\bar\theta=(\theta_1,\dots,\theta_M,\theta_1^*,\dots,\theta_M^*)^T$, and $\bar\Amat\in\mathbb{C}^{2M\times2M}$ collects all derivatives of $\bar\Wmat(\thetavec)$ as in \eqref{A_def}: 
\be \label{A_def_ausmented}
[\bar\Amat]_{n,k}\triangleq -\E\left[\sum_{m=1}^{2M} \frac{\partial  \left([\bar\Wmat(\thetavec)]_{n,m}\bar v^*_k(\thetavec)\right)}{\partial \theta_m^*}\right].
\ee
In addition, 
define
the augmented \ac{bfim} as \cite{Ollila08}
\be
\label{bar_fim}
\bf \bar\Jmat = \begin{bmatrix}
	\Jmat&
  \Pmat\\ \Pmat^* & \Jmat^*
	\end{bmatrix}\in\mathbb{C}^{2M\times2M},
 \ee
where
  $\Jmat$ is the \ac{bfim} as given in \eqref {J_def}, and the pseudo-information matrix is defined as  (cf.\ \cite{Ollila08})
\beqna \label{P_def}
\Pmat \define \E\left[\nabla_{\thetavecsmall}^T \log [p(\thetavec)p(\xvec_a|\thetavec)p(\xvec_q|\thetavec)] \right.\nonumber\\\times
\left. \nabla_{\thetavecsmall}\log [p(\thetavec)p(\xvec_a|\thetavec)p(\xvec_q|\thetavec)]\right].
\eeqna
This matrix can be decomposed into the prior, analog, and quantized contributions in a similar manner to the \ac{bfim} in \eqref{J_def}.
Finally, we define
\be
\label{bar_G_def}
\bar\Gmat
=\E\bigl[\bar\Wmat(\thetavec)\bar\Jmat\,\bar\Wmat(\thetavec)\bigr]
+\bar\Amat+\bar\Amat^H+\E[\bar\vvec(\thetavec)\,\bar\vvec^H(\thetavec)].
\ee
\begin{theorem}[Augmented WBCRB]
\label{AWBCRB}
Under the mixed‐resolution model of Subsection~\ref{math} and Conditions~\ref{cond1}–\ref{cond4}, for any estimator~$\hat\thetavec(\xvec_a,\xvec_q)$
\be
\label{augmented_bound}
\mathrm{MSE}_{\rm aug}
\;\succeq\;
\E\bigl[\bar\Wmat(\thetavec)\bigr]\;\bar\Gmat^{-1}\;
\E\bigl[\bar\Wmat(\thetavec)\bigr]^H,
\ee
where $\bar\Gmat$ is defined in \eqref{bar_G_def}.
\end{theorem}

\begin{IEEEproof}
   The proof is similar to the proof in Appendix \ref{theo_1}, with the following augmented auxiliary function:
\be
\bar\gvec(\xvec_a,\xvec_q,\thetavec)
=\bar\Wmat(\thetavec)\,\bigl[\nabla_{\rm aug}\log p\bigr]^H
\;+\;\bar\vvec(\thetavec),
\ee
where the augmented score is defined as
\be
\nabla_{\rm aug}\,\log p
=\begin{bmatrix}
\nabla_{\thetavecsmall}\log p(\thetavec,\xvec_a,\xvec_q)\\[6pt]
\nabla_{\thetavecsmall^*}\log p(\thetavec,\xvec_a,\xvec_q)
\end{bmatrix}\in{\mathbb{C}}^{2M}.
\ee
   \end{IEEEproof}
   
 The upper‐left $M\times M$ block of the bound on the r.h.s. of \eqref{augmented_bound} provides a lower bound on the ordinary \ac{mse}, while the off‐diagonal blocks bound the pseudo‐covariance of the error.  
 If the pseudo‐information in $\bar\Jmat$ (defined in \eqref{P_def}) vanishes, the augmented bound decouples into two identical real‐valued \ac{wbcrb}s on the covariance of the real and imaginary parts. 
 In this case, the upper‐left $M\times M$ block of the bound coincides with the \ac{wbcrb} from Theorem \ref{theo_1}.

As an example, it can be seen that
the inverse of ${\bf{\bar\Jmat}}$ is given by (see Equation (8) in \cite{Ollila08})
\be \label{aug_bfim} 
{\bf{\bar\Jmat}}^{-1} = \begin{bmatrix}
	{\bf{\mathcal{R}}}^{-1}&
  -{\bf{\mathcal{R}}}^{-1}{\bf{\mathcal{Q}}}\\ -{\bf{\mathcal{Q}}}^H{\bf{\mathcal{R}}}^{-1} & {\bf{\mathcal{R}}}^{{-1}^*}
	\end{bmatrix},
\ee
where $\bf{\mathcal{R}}$ is the associated  Schur complement, defined as 
$\Jmat-\Pmat (\Jmat^{-1})^*\Pmat^*$, and 
$\bf{\mathcal{Q}}\define 
\Pmat  (\Jmat^{-1})^*$. This formulation coincides with the widely linear   \ac{bcrb}, which is generally tighter than the ordinary  \ac{bcrb} \cite[p.~168]{complex_value}. 

\section{Extension of the WBCRB on the Augmented MSE Under the LGO model}
\begin{theorem}
\label{Theorem2}
Under the \ac{lgo} model and under the conditions from Section \ref{reg_subsection}, the pseudo-information matrix $\Pmat$ \eqref{P_def} vanishes. 
\end{theorem}
\begin{IEEEproof}
    The conditional log-likelihood function for the considered model is given in \eqref{joint_pdf}. 
Using a similar derivation as for $\Jmat$ in \cite{Mazor24}, it can be shown that the pseudo-information matrix $\Pmat$, defined in \eqref{P_def}, satisfies
\be	\label{pseudo_Fisher_information}
	\Pmat = \Pmat_\thetavecsmall+\E_\thetavecsmall[\Pmat_{\xvec_a|\thetavecsmall}]+\E_\thetavecsmall[\Pmat_{\xvec_q|\thetavecsmall}].
\ee 
The pseudo-information matrices based on each set of measurements $\xvec_a$ and $\xvec_q$ given $\thetavec$ are defined as: 
\begin{subequations}
\begin{align}
	\label{p_FIMa}
	&\Pmat_{\xvec_a|\thetavecsmall} \triangleq \E_{\xvec_a|\thetavecsmall}\left[\nabla_{\thetavecsmall}^T \log 
 {p}(\xvec_a|\thetavec) \nabla_{\thetavecsmall} \log  {p}(\xvec_a|\thetavec)\right],
\\	\label{p_FIMq_i}
	&\Pmat_{\xvec_q|\thetavecsmall} \triangleq \E_{\xvec_q|\thetavecsmall}\left[\nabla_{\thetavecsmall}^T\log {p}(\xvec_q|\thetavec)\nabla_{\thetavecsmall}\log {p}(\xvec_q|\thetavec)\right].
\end{align}
\end{subequations}
Furthermore,  $\thetavec$ is a circularly symmetric Gaussian vector, and
$\xvec_a|\thetavec \sim \mathcal{CN}(\Hmat\thetavec,\sigma_a^2\Imat_{N_a})$.
Thus, it follows immediately that $\Pmat_{\xvec_a|\thetavecsmall}$ and $\Pmat_\thetavecsmall =\zerovec$ vanish (note that the conditional distribution $p(\xvec_a|\thetavec)$ is a complex, circularly symmetric Gaussian).
By substituting these results in \eqref{pseudo_Fisher_information}, we obtain that for the considered model $\Pmat = \E_\thetavecsmall[\Pmat_{\xvec_q|\thetavecsmall}]$.


To complete the proof, it remains to show that $\E_\thetavecsmall[\Pmat_{\xvec_q|\thetavecsmall}]$ in \eqref{p_FIMq_i} vanishes as well. 

The 1-bit quantized log-likelihood function is equal to
\beqna \label{loglikeXq}
   \log p(\xvec_q|\thetavec) = \sum_{n=1}^{N_q} 
        \left( \frac{1}{2} + \frac{\Ree\{\xvec_{q_n}\}}{\sqrt{2}} \right) \log \Phi (\zeta_n^R ) \hspace{0.5cm}\nonumber\\ 
        + \left( \frac{1}{2} - \frac{\Ree\{\xvec_{q_n}\}}{\sqrt{2}} \right) \log \left( 1 - \Phi ( \zeta_n^R) \right) \nonumber\\ 
        + \left( \frac{1}{2} + \frac{\Imm\{\xvec_{q_n}\}}{\sqrt{2}} \right) \log \Phi (\zeta_n^I ) \hspace{1cm}
        \nonumber\\ 
        + \left( \frac{1}{2} - \frac{\Imm\{\xvec_{q_n}\}}{\sqrt{2}} \right) \log \left( 1 - \Phi (\zeta_n^I) \right),  
\eeqna
and it can be verified that the complex-valued gradient of the quantized measurements' log-likelihood from \eqref{loglikeXq} is
\beqna \label{Gradq1}
\nabla_\thetavecsmall^T \log p(\xvec_q|\thetavec) = \frac{1}{\sqrt{2}\sigma_q}\sum\nolimits_{n=1}^{N_q}\gvec_n\hspace{2cm}\nonumber\\ \times\left[
     \frac{\phi(\zeta_n^R)}{\Phi(\zeta_n^R )(\Phi(\zeta_n^R)-1)}\hspace{-0.1cm} \left(\Phi(\zeta_n^R ) - \frac{1}{2} - \frac{1}{\sqrt{2}} \Ree\{\xvec_{q_n}\} \right) \right.\nonumber\\ \left. 
      -j \frac{\phi(\zeta_n^I )}{\Phi(\zeta_n^I )(\Phi(\zeta_n^I )-1)} \hspace{-0.1cm}\left(\Phi(\zeta_n^I ) - \frac{1}{2} - \frac{1}{\sqrt{2}} \Imm\{\xvec_{q_n}\}  \right)\right]\hspace{-0.05cm},\hspace{0.05cm}
\eeqna
where $\zeta_n^R$ and $\zeta_n^I$ are defined in \eqref{zetaRI}.
Furthermore, the mean of the real and imaginary parts of $\xvec_{q_n}$ given $\thetavec$ is given by
\begin{subequations} \label{MeanXq}
\be \label{MeanReXq}
\vspace{-0.25cm}
   \sqrt{2} \E[{\text{Re}}\{\xvec_{q_n}\}|\thetavec] = 2\Phi(\zeta_n^R ) - 1,
\ee
\vspace{-0.2cm}
\be \label{MeanImXq}
    \sqrt{2}\E\left[{\text{Im}}\{\xvec_{q_n}\}|\thetavec \right] = 2\Phi(\zeta_n^I) - 1,
    \vspace{-0.15cm}
\ee
\end{subequations}
respectively. The second moment for both parts is equal to
\be \label{VarianceIMXq}
\vspace{-0.2cm}
    \E[ {\text{Re}}^2\{\xvec_{q_n}\}|\thetavec ] = \E[ {\text{Im}}^2\{\xvec_{q_n}\}|\thetavec ] = 0.5.
\ee
We note that the smoothness assumption is valid for quantized measurements since the expected value of the gradient is zero, as shown using  \eqref{MeanReXq}, \eqref{MeanImXq}, and \eqref{VarianceIMXq}.

Substituting \eqref{Gradq1} into $\Pmat_{\xvec_q|\thetavecsmall}$ and utilizing \eqref{MeanXq} and \eqref{VarianceIMXq}, along with the fact that the entries of $\xvec_q|\thetavec$ are independent and their real and imaginary parts are also independent, we obtain that
\be 
\begin{aligned}
&\Pmat_{\xvec_q|\thetavecsmall}=\\
&
\frac{1}{2\sigma_q^2}\sum\nolimits_{n=1}^{N_q}
{\E}_{\xvec_q|\thetavecsmall}\left[\left(\frac{\phi(\zeta_n^R)\left(\Phi(\zeta_n^R ) - \frac{1}{2} - \frac{1}{\sqrt{2}} \Ree\{\xvec_{q_n}\} \right)}{\Phi(\zeta_n^R)(\Phi(\zeta_n^R)-1)}\right)^2\right.
\\
&
-2j\left(\frac{\phi(\zeta_n^R)\left(\Phi(\zeta_n^R ) - \frac{1}{2} - \frac{1}{\sqrt{2}} \Ree\{\xvec_{q_n}\} \right)}{\Phi(\zeta_n^R )(\Phi(\zeta_n^R)-1)}\right)
\\
&
\left(\frac{\phi(\zeta_n^I )\left(\Phi(\zeta_n^I ) - \frac{1}{2} - \frac{1}{\sqrt{2}} \Imm\{\xvec_{q_n}\}  \right)}{\Phi(\zeta_n^I )(\Phi(\zeta_n^I )-1)}\right)
\\
&
\left.-\left(\frac{\phi(\zeta_n^I )\left(\Phi(\zeta_n^I ) - \frac{1}{2} - \frac{1}{\sqrt{2}} \Imm\{\xvec_{q_n}\}  \right)}{\Phi(\zeta_n^I )(\Phi(\zeta_n^I )-1)}\right)^2 \right]\gvec_n\gvec_n^T
\\
&
=\frac{1}{2\sigma_q^2}\sum\nolimits_{n=1}^{N_q} \left[
{\E}_{\xvec_q|\thetavecsmall}\left[\left(\frac{\phi(\zeta_n^R)\left(\Phi(\zeta_n^R )-\frac{1}{2} - \frac{1}{\sqrt{2}} \Ree\{\xvec_{q_n}\} \right)}{\Phi(\zeta_n^R)(\Phi(\zeta_n^R)-1)}\right)^2\right.\right.
\\
&
\left.-\left(\frac{\phi(\zeta_n^I )\left(\Phi(\zeta_n^I ) - \frac{1}{2} - \frac{1}{\sqrt{2}} \Imm\{\xvec_{q_n}\}  \right)}{\Phi(\zeta_n^I )(\Phi(\zeta_n^I )-1)}\right)^2 \right]
\\
&
\left.-2j{\E}_{\xvec_q|\thetavecsmall}\left[\left(\frac{\phi(\zeta_n^R)\left(\Phi(\zeta_n^R ) - \frac{1}{2} - \frac{1}{\sqrt{2}} \Ree\{\xvec_{q_n}\} \right)}{\Phi(\zeta_n^R )(\Phi(\zeta_n^R)-1)}\right)\right]\right.
\\
&
\left.{\E}_{\xvec_q|\thetavecsmall}\left[\left(\frac{\phi(\zeta_n^I )\left(\Phi(\zeta_n^I ) - \frac{1}{2} - \frac{1}{\sqrt{2}} \Imm\{\xvec_{q_n}\}  \right)}{\Phi(\zeta_n^I )(\Phi(\zeta_n^I )-1)}\right)\right]\right]\gvec_n\gvec_n^T   
\\
&
=\frac{1}{2\sigma_q^2}\sum\nolimits_{n=1}^{N_q}
\left(\frac{\phi^2(\zeta_n^R)}{\Phi(\zeta_n^R) \Phi(-\zeta_n^R)}-\frac{\phi^2(\zeta_n^I)}{\Phi(\zeta_n^I) \Phi(-\zeta_n^I)}\right) \gvec_n\gvec_n^T,
\end{aligned}
\ee
where we used the fact that
\[\E[ {\text{Re}}\{\xvec_{q_n}\}{\text{Im}}\{\xvec_{q_n}\}|\thetavec ] = \E[ {\text{Re}}\{\xvec_{q_n}\}|\thetavec ]\E[ {\text{Im}}\{\xvec_{q_n}\}|\thetavec ].\]
Thus, we have
\be
\begin{aligned}
&\E_\thetavecsmall[\Pmat_{\xvec_q|\thetavecsmall}] =
\frac{1}{2\sigma_q^2}\sum\nolimits_{n=1}^{N_q}
\left(\E_\thetavecsmall\left[\frac{\phi^2(\zeta_n^R)}{\Phi(\zeta_n^R) \Phi(-\zeta_n^R)}\right]-\right.
\\
&
\left.{\E}_\thetavecsmall\left[\frac{\phi^2(\zeta_n^I)}{\Phi(\zeta_n^I) \Phi(-\zeta_n^I)}\right]\right) \gvec_n\gvec_n^T= \zerovec.
\end{aligned}
\ee

\end{IEEEproof}
Assuming, in addition, that the off-diagonal blocks of the matrix $\Amat+\Amat^H+\E[\vvec(\thetavec)\vvec^H(\thetavec)]$ are zero, which is a condition that can be ensured by selecting an appropriate weight matrix, this claim justifies focusing solely on the l.h.s. of \eqref{bound_wbcrb}, as is widely considered in the relevant literature.  
In the context of the conventional \ac{bcrb} \eqref{bcrb}, this result further implies that the widely linear \ac{bcrb} from \cite[eq. 6.64]{complex_value}, which generally offers a tighter bound than the standard \ac{bcrb}, reduces to the traditional form and becomes equal to the \ac{bcrb}, as formalized in Corollary 1 of \cite{Ollila08}.

\end{document}

%% file: myshorts.tex
\newcommand{\avec}{{\bf{a}}}

\newcommand{\bvec}{{\bf{b}}}
\newcommand{\cvec}{{\bf{c}}}

\newcommand{\evec}{{\bf{e}}}
\newcommand{\fvec}{{\bf{f}}}

\newcommand{\uvec}{{\bf{u}}}
\newcommand{\wvec}{{\bf{w}}}
\newcommand{\xvec}{{\bf{x}}}

\newcommand{\tvec}{{\bf{t}}}

\newcommand{\vvec}{{\bf{v}}}
\newcommand{\gvec}{{\bf{g}}}

\newcommand{\onevec}{{\bf{1}}}
\newcommand{\zerovec}{{\bf{0}}}

\newcommand{\alphavec}{{\bf{\alpha}}}
\newcommand{\epsilonvec}{{\boldmath{\epsilon}}}

\newcommand{\Amat}{{\bf{A}}}

\newcommand{\Cmat}{{\bf{C}}}
\newcommand{\Dmat}{{\bf{D}}}

\newcommand{\Gmat}{{\bf{G}}}
\newcommand{\Hmat}{{\bf{H}}}
\newcommand{\Jmat}{{\bf{J}}}
\newcommand{\Imat}{{\bf{I}}}

\newcommand{\Kmat}{{\bf{K}}}
\newcommand{\Pmat}{{\bf{P}}}

\newcommand{\Tmat}{{\bf{T}}}

\newcommand{\Wmat}{{\bf{W}}}

\newcommand{\Zmat}{{\bf{Z}}}

\newcommand{\Imm}{{\rm Im}}
\newcommand{\Ree}{{\rm Re}}

\newcommand{\define}{\stackrel{\triangle}{=}}

\newcommand{\E}{{\mathbb{E}}}

\def\btau{{\mbox{\boldmath $\tau$}}}

\def\Sigmavec{{\mbox{\boldmath $\Sigma$}}}

\def\btau{{\mbox{\boldmath $\tau$}}}

\def\betavec{{\mbox{\boldmath $\beta$}}}

\def\alphavec{{\mbox{\boldmath $\alpha$}}}

\def\thetavec{{\mbox{\boldmath $\theta$}}}

\def\muvec{{\mbox{\boldmath $\mu$}}}

\def\thetavecsmall{{\mbox{\boldmath {\scriptsize $\theta$}}}}

\newcommand{\be}{\begin{equation}}
\newcommand{\ee}{\end{equation}}
\newcommand{\beqna}{\begin{eqnarray}}
\newcommand{\eeqna}{\end{eqnarray}}
